\documentclass[pdflatex,sn-nature]{sn-jnl}
\usepackage{graphicx}
\usepackage{amsmath,amssymb,amsfonts}
\usepackage{textcomp}
\usepackage{booktabs,longtable,array,calc,tabularx}
\newsavebox{\tbox}
\DeclareUnicodeCharacter{2295}{\ensuremath{\oplus}}
\DeclareUnicodeCharacter{2016}{\ensuremath{\Vert}}
\DeclareUnicodeCharacter{03A9}{\ensuremath{\Omega}}
\DeclareUnicodeCharacter{2020}{\textdagger}
\usepackage{xcolor}
\usepackage{placeins}
\usepackage{etoolbox}
\DeclareUnicodeCharacter{2212}{\ensuremath{-}}
\DeclareUnicodeCharacter{2265}{\ensuremath{\geq}}
\DeclareUnicodeCharacter{2264}{\ensuremath{\leq}}
\DeclareUnicodeCharacter{2248}{\ensuremath{\approx}}
\DeclareUnicodeCharacter{2260}{\ensuremath{\neq}}
\DeclareUnicodeCharacter{221A}{\ensuremath{\surd}}
\DeclareUnicodeCharacter{0394}{\ensuremath{\Delta}}
\DeclareUnicodeCharacter{03A3}{\ensuremath{\Sigma}}
\DeclareUnicodeCharacter{03B2}{\ensuremath{\beta}}
\DeclareUnicodeCharacter{03B3}{\ensuremath{\gamma}}
\DeclareUnicodeCharacter{03BB}{\ensuremath{\lambda}}
\DeclareUnicodeCharacter{03BC}{\ensuremath{\mu}}
\DeclareUnicodeCharacter{03C1}{\ensuremath{\rho}}
\DeclareUnicodeCharacter{03C3}{\ensuremath{\sigma}}
\DeclareUnicodeCharacter{2192}{\ensuremath{\rightarrow}}

\unnumbered
\makeatletter
\patchcmd{\@maketitle}{Corresponding author(s). E-mail(s): }{Corresponding author. E-mail: }{\typeout{PATCHOK title}}{\typeout{PATCHFAIL title}}
\patchcmd{\email}{\textcolor{blue}{#1};\ }{\textcolor{blue}{#1}}{\typeout{PATCHOK email}}{\typeout{PATCHFAIL email}}
\makeatother
\begin{document}
\title[Shared clones make random cell halves look reproducible]{Shared clones make random cell halves look reproducible in pooled screens}
\author*{\fnm{Xuanrui} \sur{Huang}}\email{huangxuanrui@u.nus.edu}
\affil{\orgdiv{Department of Biomedical Informatics, Yong Loo Lin School of Medicine}, \orgname{National University of Singapore}, \orgaddress{\country{Singapore}}}
\abstract{Single-cell perturbation screens are used to train and evaluate models that predict the effects of perturbations. A model's score is compared with the ceiling that the data themselves allow. This ceiling, and which perturbations have reliable effects, are usually judged by splitting each perturbation's cells into two random halves and asking whether the effects measured in the two halves point the same way. This assumes that the halves are independent replicates. In pooled single-cell screens such as Perturb-seq, however, guide RNAs are delivered by lentivirus at a low dose and the cells then expand, so the cells that carry a guide descend from a limited number of transduced cells. All descendants of one transduced cell form a clone, and two random halves contain the same clones. This is pseudoreplication, and its unit, the clone, is not marked in the data. Using perturbations that should have no effect as controls, we compared random halves with independent replicates from separate
transductions, and tested directly with halves that differ only in their clones. In a genome-scale screen of primary T cells, perturbations that should have no effect pointed the same way between random halves almost always and between independent replicates only half the time, the rate expected from noise alone; measuring more cells did not remove this agreement. When the cells were split into halves that share no clone, by the constant region of the T-cell receptor or by a clone barcode in the vector, the excess agreement largely disappeared. Under a simple model, consistent with a screen that recorded clones, the size of this excess equals the part of the expression differences between cells that is set by clones, divided by the effective number of clones per perturbation; the effective number is smaller when fewer cells are transduced per guide and when their expansion is more uneven. In cell-line screens, an upper bound on this excess was small, and in their CRISPRi screens also
small relative to the real effects. In the T-cell screen, shared clones inflated benchmark ceilings and reliability labels, the more so in donors whose perturbations had fewer effective clones. In this screen, whose cell states came from one transduction, when a model's input and the answer used to score it come from cells of the same transduction, published models that transfer perturbation effects to a new cell state scored well even on perturbations that should have no effect; their excess over scores against the second guide of the same gene came from shared clones, since for STATE, the published model tested with the same guide from another transduction as the answer, the gain disappeared. Reproducibility is better judged with cells from separate transductions, which share no clone, than with random halves; where none are available, null perturbations show how far random halves can be trusted.}
\keywords{Perturb-seq, pooled CRISPR screens, reproducibility, pseudoreplication, clonal expansion, benchmarking}
\maketitle

\section{Introduction}

\begin{figure}[!tbp]
\sbox{\tbox}{\begin{minipage}{\textwidth}\makebox[\textwidth][c]{\includegraphics[width=170mm]{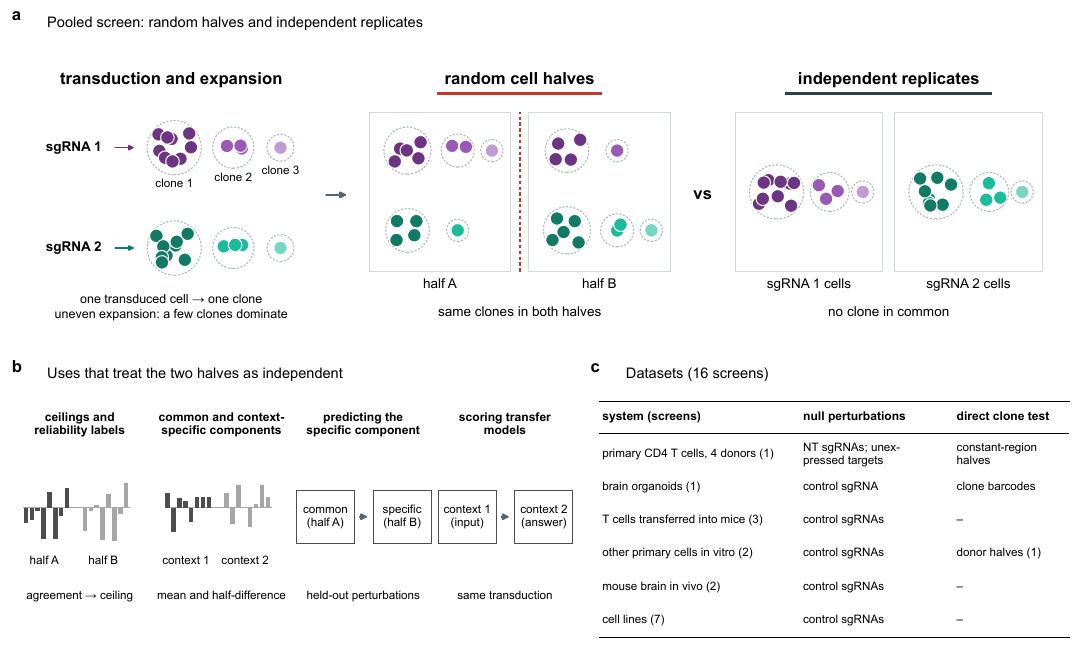}}
\caption{\textbf{Random halves contain the same clones; independent replicates do not.} \textbf{a}, Why random halves are not independent replicates. Left, a perturbation is targeted by two sgRNAs that enter different cells (sgRNA 1, purple; sgRNA 2, green); the cells carrying one sgRNA descend from a few transduced cells, and all descendants of one transduced cell form a clone (dashed rings; shades of one colour distinguish clones). Expansion is uneven, so a few clones contribute most of the cells (clones 1--3 have 9, 3 and 1 cells). Middle, when all cells of the perturbation are split at random into two halves, every clone with several cells is present in both halves, so the halves contain the same clones (red line). Right, the cells of the two sgRNAs come from two transductions and share no clone (dark blue line). \textbf{b}, Four practices that treat two halves of one transduction as independent samples. From left: benchmark ceilings and reliability labels, from the agreement between two
half-effects corrected by the Spearman--Brown formula; common and context-specific parts, the mean and the half-difference of a perturbation's effects in two contexts (such as resting and stimulated); predicting the specific part, in which a predictor learns on some perturbations how the specific part relates to the common part and is scored on the others, with the common part taken from one half of the cells (A) and the specific part from the other (B) so that they share no sampling noise; and scoring transfer models, which predict the effect in context 2 from the cells in context 1, with an answer from the same transduction as the input. \textbf{c}, The 16 screens analysed, grouped by system, with the number of screens in parentheses. Null perturbations are perturbations that should have no effect: non-targeting (NT) sgRNAs, or sgRNAs against genes not expressed in the cells. A direct clone test splits the cells so that the halves carry the same sgRNA but share no clone: by the TCRβ
constant region (TRBC) in CD4, and by the clone barcode in the vector in the organoids; in one primary T-cell screen, halves from two donors carry the same sgRNAs but share no clone (donor halves); the other screens have no such split.}\label{fig:1}\end{minipage}}
\typeout{FIG 1 \the\dimexpr\ht\tbox+\dp\tbox\relax\space\the\textheight}\usebox{\tbox}
\end{figure}

Single-cell perturbation screens now measure the transcriptional changes caused by thousands of genetic perturbations across cell types and cell states\textsuperscript{1--3}. These data are used to train and evaluate so-called virtual-cell models, which predict the effects of perturbations or cellular contexts that have not been measured\textsuperscript{4,5}. How good a model's score is has to be judged against the ceiling that the data themselves allow: only the part of a measured effect that reappears in a repeated experiment can be predicted. Recent work has therefore argued that benchmarks should report the reproducibility ceiling of the data alongside model scores\textsuperscript{6}.

The usual way to estimate reproducibility is to split each perturbation's cells into two random halves and compare the effects measured in the two; we call this random halves. At least three practices rest on it. The evaluation tool cell-eval uses the agreement between random halves, corrected by the Spearman--Brown formula, as the ceiling of the data, and its documentation describes this as an unbiased upper bound on how well any model could score\textsuperscript{7--9}. Per-perturbation reliability labels, proposed for cleaning training and evaluation data, are based on the same agreement\textsuperscript{6}. And a class of methods splits perturbation effects into a part common to all cellular contexts and a part specific to each context, uses random halves to separate reproducible variation from noise, and builds on this split to transfer effects to new contexts\textsuperscript{10}. All three assume that the cells of a perturbation are independent samples, so that two subsets of them
differ only by sampling noise. The evaluation of transfer models makes the same assumption implicitly: a perturbation's cells in the source context are used for training and its cells in the target context serve as the answer for scoring\textsuperscript{11}; if the cells of the two contexts come from the same transduction, they are not independent either.

In a widely used class of screens this assumption does not hold. These screens have three features. First, guide RNAs (sgRNAs) are introduced into cells by an integrating lentivirus, a step called transduction; to give each cell a single sgRNA, the dose of virus is low, and the sgRNA passes to the cell's descendants with its genome. Second, the cells divide and expand after transduction, for example during drug selection, culture, proliferation after transfer into an animal, or the development of an organoid. Third, single-cell sequencing reads each cell's sgRNA together with its transcriptome, and cells are then grouped by sgRNA (Perturb-seq and related methods). The cells that carry a given sgRNA at sequencing therefore all descend from a limited number of transduced cells, whose expansion is often very uneven\textsuperscript{12--14}. All descendants of one transduced cell form a clone, and a clone has its own expression state. Random halves come from the same transduction and may
contain the same clones, so their agreement may reflect the state of those clones rather than the effect of the perturbation. We say that two sets of cells come from the same transduction when they descend from the same transduced cells, and from separate transductions when they descend from different transduced cells, even if all cells were infected in one step. Two sets of cells from separate transductions, such as the cells of the two sgRNAs against one gene or the cells of different donors, share no clone; we call them independent replicates.

Statistically this is pseudoreplication: observations that are not independent are treated as independent replicates. Single-cell differential expression analysis has long recognized that treating cells from the same individual as independent samples inflates false positives\textsuperscript{15,16}. In screens at high multiplicity of infection, where each cell carries many sgRNAs, the combination of sgRNAs identifies clones, and clonal expansion has been shown to produce false-positive differentially expressed genes\textsuperscript{17}. It has also been noted that random halves measure sampling precision rather than experimental reproducibility\textsuperscript{18}, and a concurrent study shows that when cells of one clone or patient fall on both sides of a random split, supervised models can score well by retrieving a relative of the test cell\textsuperscript{19}. In the common screens at low multiplicity of infection, however, clones carry no label in the data. Individuals are labelled
and can be pooled or modelled; invisible clones cannot be handled in this way. How large this effect is in such screens, what determines its size and what it does to the practices above has, to our knowledge, not been measured.

Here we measure it in screens that contain independent replicates (Fig. 1). We compare random halves with independent replicates, using perturbations that should have no effect as controls, and test directly under conditions in which the two halves differ only in their clones: once with a natural marker of T cells and once with a clone barcode in the vector. We then measure the size of the effect in 16 screens and relate it to the effective number of clones per perturbation. Finally, we test what it does to benchmark ceilings, reliability labels, cross-context structure and the evaluation of transfer models.

\section{Results}

\subsection{Null perturbations look reproducible under random halves}

\begin{figure}[!tbp]
\sbox{\tbox}{\begin{minipage}{\textwidth}\makebox[\textwidth][c]{\includegraphics[width=170mm]{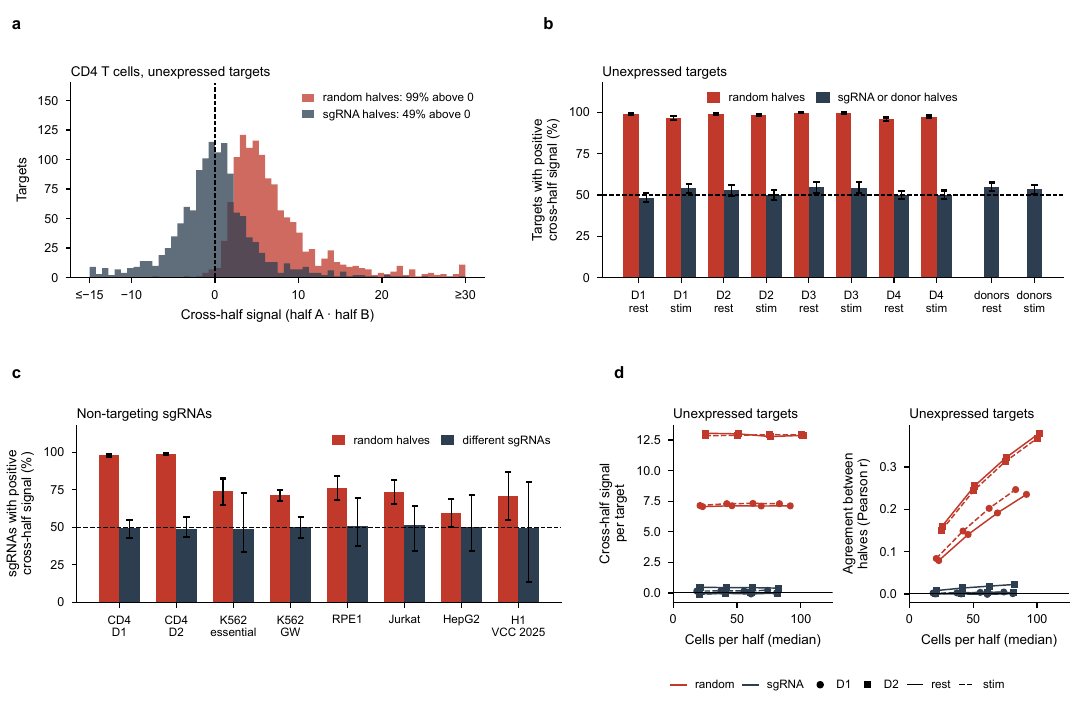}}
\caption{\textbf{Null perturbations look reproducible under random halves, and more cells do not remove this.} \textbf{a}, Do the two halves of a perturbation with no expected effect agree? Distribution of the cross-half signal of unexpressed targets in resting CD4+ T cells of donor 1 (n = 1,239 targets). In each half, the perturbation's cells were compared with control cells, which gives an expression difference per gene (log1p CP10K); the cross-half signal is the gene-wise product of the two halves' differences, summed over genes, and is positive or negative with equal probability when the halves differ only by sampling noise. Random halves (red) and sgRNA halves (dark blue; the halves are the cells of the two sgRNAs): 99\% positive and shifted to the right against 49\%, as with noise alone. Values beyond −15 and 30 are placed in the end bins. \textbf{b}, Does this hold in all four donors, both states and another kind of independent replicate? Percentage of unexpressed targets with a positive
cross-half signal. The first eight groups are single donors, resting (rest) and stimulated (stim): random halves (red) and sgRNA halves (dark blue); n = 1,235, 869, 807 and 1,459 targets in donors 1--4, those with at least 10 cells on each side of every halving in both states. The last two groups are donor halves, the cells of donors 1 and 2 against those of donors 3 and 4, which were transduced separately and share no clone (n = 1,479). Dashed line, 50\%, the chance level. Error bars, bootstrap 95\% intervals over targets (1,000 resamples). \textbf{c}, Is it the same for another kind of null perturbation and in cell lines? Percentage of non-targeting sgRNAs with a positive cross-half signal. Red, random halves of one non-targeting sgRNA; dark blue, half of the cells of one non-targeting sgRNA against half of another's, which share no clone (mean over 50 random pairings; error bars, range). CD4 donors 1 and 2 (n = 739 and 578 sgRNAs), five CRISPRi screens of cell lines (K562 essential,
97; K562 genome-wide (GW), 507; RPE1, 113; Jurkat, 128; HepG2, 112) and the H1 data of the 2025 Virtual Cell Challenge (31). Red error bars, bootstrap 95\% intervals over sgRNAs. In cell lines the share under random halves is also above 50\%, but lower than in CD4.}\label{fig:2}\end{minipage}}
\typeout{FIG 2 \the\dimexpr\ht\tbox+\dp\tbox\relax\space\the\textheight}\usebox{\tbox}
\end{figure}
\begin{figure}[!tbp]
{\small\textbf{Figure 2 (continued).} \textbf{d}, Does the excess agreement disappear with more cells? Cross-half signal (left) and between-half Pearson correlation (right) of unexpressed targets when each half uses 1/8, 1/4, 3/8 or 1/2 of a perturbation's cells (horizontal axis, median cells per half). Random splits (red), in which both halves contain cells of both sgRNAs; sgRNA splits (dark blue), in which one half comes from each sgRNA with the same number of cells as under random splits. Each point is the mean of several disjoint splits, and all cell numbers use the same targets. Circles, donor 1 (n = 1,178 resting and 1,151 stimulated targets); squares, donor 2 (709 and 725); solid lines, resting; dashed lines, stimulated. Under random splits the cross-half signal does not change with the number of cells while the correlation rises; under sgRNA splits both are close to zero. The cross-half signal is computed after subtracting the mean effect of the unexpressed targets; the correlation uses the effects relative to
control cells.\par}
\end{figure}

We first asked whether a perturbation with no expected effect looks reproducible when its cells are split into random halves. The genome-scale CRISPR interference (CRISPRi) screen of primary human CD4+ T cells\textsuperscript{3} profiles cells of four donors in a resting and a stimulated state, with two sgRNAs against each gene. We used two kinds of perturbation with no expected effect, together called null perturbations. The first are sgRNAs against genes not expressed in T cells. The screen measures expression with probes against individual genes, so the expression of each target gene is itself measured, and in each donor 807--1,459 of the targets with enough cells are not expressed in T cells (Methods); an sgRNA against a gene that is not expressed cannot change the transcriptome through that gene. The second are non-targeting control sgRNAs, which have no target in the genome; these control cells are also the cells that benchmark tools split at random when they compute ceilings.

Agreement was computed as follows. For a perturbation, its cells in each half were compared with control cells, which gives an expression difference for each gene; together these form the effect in that half. The effects of the two halves were multiplied gene by gene and summed over genes, which gives the cross-half signal. A positive value means that the two effects point broadly the same way, and its expected value equals the energy, the sum of squares over genes, of the part of the effect that the two halves share. If the halves differ only by sampling noise, the two effects fluctuate around zero independently, the cross-half signal is positive or negative with equal probability, and about half of the null perturbations are positive; our pipeline returns this in simulated screens (Supplementary Fig. S1a).

Under random halves, 96--100\% of the unexpressed targets had a positive cross-half signal, in all four donors and both states (Fig. 2a,b). With independent replicates the share fell to near chance: 48--55\% with one sgRNA in each half, and 53--55\%, slightly above one half, with donors 1--2 against donors 3--4 (Fig. 2b). In CD4, non-targeting sgRNAs behaved in the same way: 97--99\% were positive under random halves, against 49--50\% for halves taken from two different non-targeting sgRNAs; in cell lines, the share under random halves was also above one half (60--76\%) but lower than in CD4 (Fig. 2c; Supplementary Table S1). These shares did not depend on the choice of genes: with the 2,000 genes of highest control mean, with genes of control mean at least 0.5, or without mitochondrial and ribosomal-protein genes, 99--100\% of the unexpressed targets in resting cells of donors 1 and 2 were positive under random halves and 49--52\% under sgRNA halves (Methods). Random halves therefore
report a reproducible response for perturbations that should have none, and independent replicates do not. We call the excess agreement of random halves over independent replicates the shared component.

The halves of an independent replicate differ not only in their clones but also in the sgRNA or the donor, so these results do not yet show that the shared component comes from clones. If each sgRNA had an action of its own beyond its target, both random halves would contain cells of that sgRNA and would also agree. The next two sections first exclude sampling noise and then answer this question with halves that carry the same sgRNA and differ only in their clones.

\subsection{More cells do not remove the shared component}

If the shared component were sampling noise, measuring more cells would shrink it. To test this, we split each perturbation's cells at random into eight parts and let each half use 1/8, 1/4, 3/8 and 1/2 of the cells in turn, on the same targets (Fig. 2d). Under random splits, the cross-half signal of null perturbations was the same at all four cell numbers (7.05--7.12 in resting cells of donor 1, from 23 to 92 cells per half); with the two sgRNAs on opposite sides, it was close to zero at every cell number. The Pearson correlation between the two halves of these null perturbations nevertheless rose with the number of cells, from 0.08 to 0.24 in donor 1. This is what a shared component that does not change with cell number predicts: with more cells, sampling noise falls while the shared component stays, so the correlation between random halves keeps rising.

This has a direct consequence for benchmark ceilings. Benchmarks extrapolate the split-half correlation to all cells with the Spearman--Brown formula and use the result as the ceiling. The formula predicted the rise above correctly (from the correlation at 1/4 of the cells it predicts 0.25 at 1/2, against an observed 0.24), but what it extrapolates is the agreement within the same clones: however many cells are measured, the ``ceiling'' of a null perturbation approaches the agreement within its clones, not zero.

\subsection{The halves share clones}

\begin{figure}[!tbp]
\sbox{\tbox}{\begin{minipage}{\textwidth}\makebox[\textwidth][c]{\includegraphics[width=170mm]{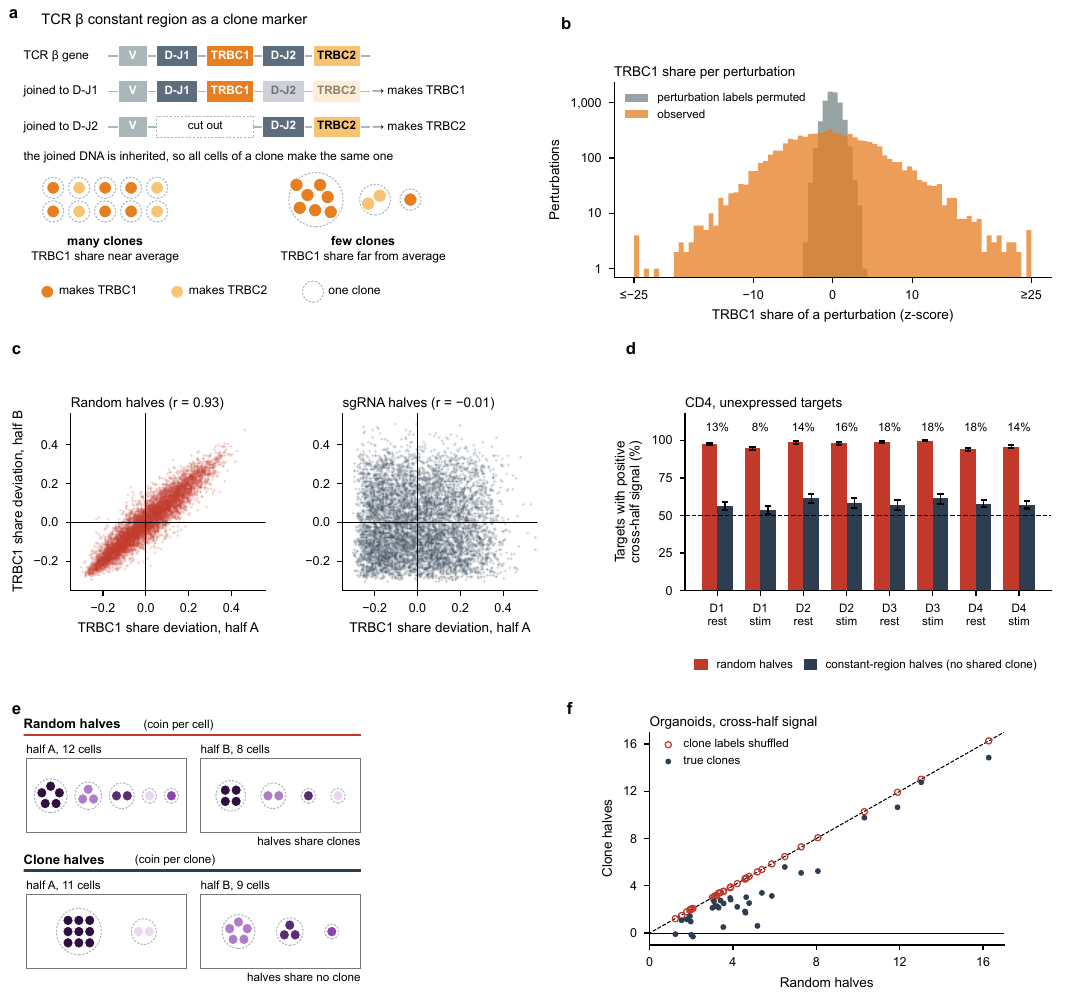}}
\caption{\textbf{The halves share clones.} \textbf{a}, The TCRβ constant region as a clone marker. The TCRβ gene contains, in order, V, D-J1, TRBC1, D-J2 and TRBC2. When V is joined to D-J1, the transcript is spliced to the nearest constant region and the cell makes TRBC1; when V is joined to D-J2, D-J1 and TRBC1 between them are cut out and the cell can make only TRBC2. The joined DNA is inherited, so the cells of one clone use the same constant region. A perturbation whose cells come from many clones has a TRBC1 share near the population average; one whose cells come from few clones deviates from it (bottom).}\label{fig:3}\end{minipage}}
\typeout{FIG 3 \the\dimexpr\ht\tbox+\dp\tbox\relax\space\the\textheight}\usebox{\tbox}
\end{figure}
\begin{figure}[!tbp]
{\small\textbf{Figure 3 (continued).} \textbf{b}, How far each perturbation's TRBC1 share deviates from the population average. For each perturbation, the mean TRBC1 share of its cells with TRBC counts (at least 30) minus the population mean was divided by the standard error expected for randomly drawn cells (per-cell standard deviation/√cells), which gives a z-score. If a perturbation's cells came from many clones, z would be close to standard normal, with variance 1, as it is after shuffling perturbation labels among cells (grey). Observed (orange), the variance is 36: each perturbation's cells come from a few clones. Resting cells of donor 1, n = 6,610 perturbations; logarithmic vertical axis; values beyond ±25 are placed in the end bins. \textbf{c}, Each point is a perturbation. Its cells are split into two halves, and the TRBC1 deviation of a half is its mean TRBC1 share minus the population mean; horizontal axis, half A; vertical axis, half B. Left, random halves, which contain the same clones: the deviations are
nearly equal (r = 0.93). Right, sgRNA halves, which come from two transductions and share no clone: the deviations are unrelated (r = −0.01). Resting cells of donor 1, at least 15 cells with TRBC counts per half; n = 6,568 (left) and 6,072 (right) perturbations. \textbf{d}, Constant-region halves: one half holds a perturbation's cells in which only TRBC2 was detected (at least 4 TRBC counts) and the other its cells with a TRBC1 share of at least 0.5 (at least 2 counts); the halves have the same sgRNA, donor and state but share no clone. The comparison is random halves of the same classified cells. Percentage of unexpressed targets with a positive cross-half signal under random halves (red) and constant-region halves (dark blue). Numbers above bars, the cross-half signal kept by constant-region halves as a percentage of that of random halves. n = 1,581 resting and 1,502 stimulated targets (donor 1), 968 and 981 (donor 2), 835 and 768 (donor 3), 1,889 and 1,673 (donor 4). Error bars,
bootstrap 95\% intervals over targets; dashed line, 50\%; paired sign test of random minus constant-region halves, p \textless{} 10\textsuperscript{−180} in each group. \textbf{e}, Schematic. A perturbation's 20 cells come from 5 clones of 9, 5, 3, 2 and 1 cells (one shade of purple per clone; dashed rings). Random halves (red line) toss a coin for each cell: the halves receive 12 and 8 cells, and every clone with several cells is present in both. Clone halves (dark blue line) toss a coin for each clone, so that each clone goes whole to one half: the halves receive 11 and 9 cells and share no clone; their cell numbers therefore need not be equal. \textbf{f}, The vectors of the human brain organoid screen carry clone barcodes, so each cell's clone is known. Each point is a perturbation (n = 32): horizontal axis, cross-half signal of random halves; vertical axis, cross-half signal of clone halves (dark blue), each the mean of 1,000 splits. Both kinds of halves carry the same sgRNA and
differ only in whether they share clones; if clones played no role, the points would lie on the diagonal (dashed). All 32 lie below it (sign test, p = 2 × 10\textsuperscript{−10}), 31\% lower in total; what remains is the part not shared through clones. Open red points, the permutation control: clone labels shuffled among the cells of each perturbation before clone halves are formed (mean of 200 permutations); the points return to the diagonal. Over the 32 perturbations the mean cross-half signal is 4.98 with shuffled labels, the same as under random halves (4.97), and 3.43 with the true clones; the observed difference exceeds all 200 permutations (permutation test, p = 0.005, the smallest value possible with 200 permutations).\par}
\end{figure}

If the shared component comes from clones, a quantity that marks clones should agree between random halves and not between the two sgRNAs. T cells carry such a marker. During its development, each T cell cuts its T-cell receptor (TCR) β gene and joins it at one of two D-J segments. Joining at the second cuts out the first segment together with the TRBC1 constant region, so the cell can make only TRBC2; joining at the first makes the cell produce TRBC1\textsuperscript{20}. The joined DNA passes to all of the cell's descendants, so the cells of one clone use the same constant region (Fig. 3a), and the readout of the CD4 screen measures both constant regions. If a perturbation's cells came from many clones, their TRBC1 share would be close to that of the whole population; if they came from few, it would deviate from it. Scaled as a z-score against random assignment of cells, the deviation would have variance 1 in the first case; it was 36 for the perturbations of the main analysis in
resting cells of donor 1 (Fig. 3b), and between 17 and 81 over all perturbations of the four donors and two states (34 for resting cells of donor 1; Supplementary Table S2a). The deviation was nearly the same between the random halves of a perturbation (r = 0.93 in donor 1, 0.86--0.97 in all donors and states) and unrelated between its two sgRNAs (\textbar r\textbar{} ≤ 0.02; Fig. 3c). Random halves contain the same clones, and the two sgRNAs contain different ones.

The marker also allows a direct test. When a perturbation's cells are split by constant region, one half holding cells in which only TRBC2 was detected and the other cells with a high TRBC1 share, the halves share no clone, while the sgRNA, donor, library and state are the same. Each cell was referenced to control cells of the same constant-region class, and the TCR genes themselves were excluded (Methods). For comparison, the same cells were split at random. In resting and stimulated cells of all four donors, 94--100\% of the unexpressed targets were positive under random halves and 54--61\% under constant-region halves, near the 50\% expected by chance though above it; the cross-half signal kept only 8--18\% of its random-half value, and 89--98\% of the targets had a higher signal under random halves (paired sign test, p \textless{} 10\textsuperscript{−180} in each donor and state; Fig. 3d). The halves carry the same sgRNA, so an action of the sgRNA itself cannot explain the
difference. The small remainder may come from clones leaking between the halves through the few cells assigned to the wrong class, or from a component shared by all perturbations; this was not tested.

TRBC distinguishes only two classes of clones. A screen in human brain organoids\textsuperscript{21} added a unique clone barcode to the vector at each integration and sequenced it together with each cell's transcriptome, which allows a test clone by clone. From these barcodes we assigned clones to 10,967 cells (Methods) and split each perturbation's cells by clone: all cells of a clone went to the same half, and the two halves carried the same sgRNA and came from the same organoid libraries (Fig. 3e). In all 32 perturbations the cross-half signal was higher under random halves than under clone halves (sign test, p = 2 × 10\textsuperscript{−10}; Fig. 3f), and 31\% of the random-half signal was present only within the same clones. When the clone labels were shuffled among the cells of each perturbation, the difference disappeared (permutation test, p = 0.005, the smallest value possible with 200 permutations). The two tests, in a system with a strong clone effect and in one with a weak
clone effect, reach the same conclusion: the shared component comes from clones, not from the sgRNA.

\subsection{The size of the shared component depends on the effective number of clones per perturbation}

\begin{figure}[!tbp]
\sbox{\tbox}{\begin{minipage}{\textwidth}\makebox[\textwidth][c]{\includegraphics[width=170mm]{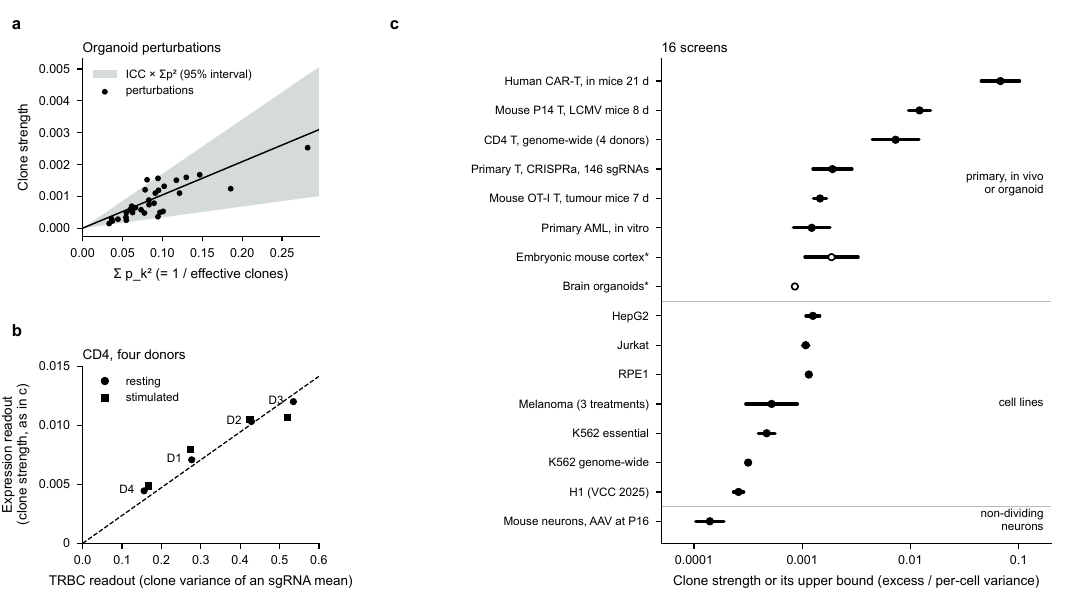}}
\caption{\textbf{The size of the shared component depends on the effective number of clones per perturbation.} Clone strength, the excess agreement that two halves owe to shared clones in units of the per-cell variance, equals ICC × Σ p\textsubscript{k}\textsuperscript{2}, where p\textsubscript{k} is the share of clone k among a perturbation's cells, Σ p\textsubscript{k}\textsuperscript{2} is the inverse of the effective number of clones and the ICC is the share of the expression differences between cells that is set by clones. \textbf{a}, Brain organoid screen; each point is a perturbation (n = 32). Horizontal axis, Σ p\textsubscript{k}\textsuperscript{2} from the clone barcodes; vertical axis, clone strength, the difference between the cross-half signals of random halves and clone halves divided by the per-cell variance (each the mean of 1,000 splits). Line, the prediction ICC × Σ p\textsubscript{k}\textsuperscript{2} with the ICC (median 0.010) measured in the same cells with a different
estimator (analysis of variance), not fitted to these points; grey band, the 95\% interval of the ICC (2,000 bootstrap resamples of perturbations). \textbf{b}, The CD4 screen has no clone barcodes; this panel tests a consequence of the same relation. Each perturbation has two clone readouts that are both proportional to Σ p\textsubscript{k}\textsuperscript{2}: the TRBC readout (horizontal axis), the variance of each sgRNA's mean TRBC1 share beyond sampling, in units of the per-cell variance; and the expression readout (vertical axis), the clone strength given by the estimator of panel c (geometric mean of the two references). Their ratio equals the ratio of their ICCs and does not depend on the effective number of clones, so the points of all donors should lie near one line through the origin (dashed, least squares through the origin). Four donors (D1--D4), resting (circles) and stimulated (squares): each readout differs about threefold between donors, while their ratio stays between
0.021 and 0.029, falling slightly as the readouts rise, which the relation does not predict (Supplementary Table S2b).}\label{fig:4}\end{minipage}}
\typeout{FIG 4 \the\dimexpr\ht\tbox+\dp\tbox\relax\space\the\textheight}\usebox{\tbox}
\end{figure}
\begin{figure}[!tbp]
{\small\textbf{Figure 4 (continued).} \textbf{c}, Clone strength in 16 screens (logarithmic axis), with one estimator: the random-half cross-half signal of null sgRNAs, minus the cross-half signal between two different null sgRNAs (which share no clone, so the difference removes the common shift introduced by the reference), divided by the per-cell variance. The difference also contains any effect specific to each null sgRNA, so outside the CD4 screen and the organoids the values are upper bounds on clone strength. Screens are grouped into three systems separated by lines: primary, in vivo or organoid; cell lines; non-dividing neurons. Effects are referenced to the null cells or to all cells of the same sample; lines show the range over both references and over states or donors, and points their geometric midpoints. The two screens marked with an asterisk have a single null sgRNA and were estimated with clone halves (organoids) or library halves (embryonic cortex) instead (open points). Numbers of null sgRNAs in
Supplementary Table S3. The systems differ by nearly three orders of magnitude.\par}
\end{figure}

How large the shared component is follows from its source. Suppose that a perturbation's cells come from several clones, that clone k makes up a share p\textsubscript{k} of all the cells and that its expression deviates by c\textsubscript{k}. Two random halves each contain about the same share of every clone, so the deviation they have in common is Σ p\textsubscript{k}c\textsubscript{k}, whose expected energy is proportional to Σ p\textsubscript{k}\textsuperscript{2}; sampling noise of different cells is uncorrelated between the halves and does not enter this term (derivation in Supplementary Note 1).

Dividing the energy of this common deviation by the per-cell variance gives a quantity that does not change with the number of cells, which we call clone strength. It can be measured directly with null perturbations: the cross-half signal of random halves, minus the cross-half signal between two different null perturbations, divided by the per-cell variance. The derivation shows that clone strength is the ratio of two factors. The numerator is the intraclass correlation (ICC), the share of the expression differences between single cells that is set by clones; it depends on the cell type. The denominator is the effective number of clones, K\textsubscript{eff} = 1/Σ p\textsubscript{k}\textsuperscript{2}: the number of clones when all are equally large, and smaller the more their sizes differ. If N\textsubscript{0} cells were transduced per sgRNA and the clone sizes after expansion have coefficient of variation CV, then approximately K\textsubscript{eff} ≈ N\textsubscript{0}/(1 +
CV\textsuperscript{2}). This is a special case of the design effect of cluster sampling\textsuperscript{22--24}: the effective sample size is set by the distribution of cluster sizes, not by the number of individuals.

In the organoid screen, the clone composition and the ICC of each perturbation can be measured directly, so this relation can be checked there. Each perturbation had a median of 51 clones of uneven size, and a median K\textsubscript{eff} of 12; the ICC measured by analysis of variance had a median of 0.010, so within a cell type, clones set only about 1\% of the expression differences between cells. Across the 32 perturbations, clone strength increased with Σ p\textsubscript{k}\textsuperscript{2} (Spearman correlation 0.80) with a proportionality constant of 0.0096 (95\% interval from a bootstrap over perturbations, 0.0085--0.0109), in agreement with the ICC of 0.010 (0.003--0.017; Fig. 4a) measured in the same cells with a different estimator, analysis of variance. When the ICC measured in one half of the perturbations was used to predict the clone strength of the other half, the measured total was 0.94 times the predicted total (median over 500 random splits, interquartile range
0.71--1.39; over all perturbations, the bootstrap interval of the ratio of the slope to the ICC is 0.57--2.70). Both quantities come from the same cells and clone labels, so this is a check that the clone-half difference behaves as the model predicts, not an independent measurement. The CD4 screen has no clone-by-clone labels, but it allows an indirect test: each perturbation has two clone readouts, the deviation of its TRBC1 share and the shared component of its expression. Under this relation both scale with the same K\textsubscript{eff}, so their ratio should stay the same across donors. The two readouts each differed about threefold among the four donors, whereas their ratio stayed between 0.021 and 0.029, falling slightly as the readouts rose, which the relation does not predict (Fig. 4b; Supplementary Table S2b).

A small effective number of clones can come from few transduced cells or from uneven expansion; the first is set by the design and the second by the cells and their culture, and the remedies differ. In CD4 the design value can be compared with the effective number. The size of the TRBC1 deviation implies that a typical perturbation's cells behave as if they came from about 7, 4, 3 and 12 or fewer equally sized clones in donors 1--4 (medians of per-perturbation bounds; a pooled estimate gives 5, 2, 2 and 9; Methods), whereas the design transduced of the order of 500 cells per perturbation. Many cells were transduced and yet the effective number of clones is about ten or fewer, so in CD4 it is set after transduction: by very uneven expansion, in which a few clones expanded far more than the rest, or by losses of transduced cells, which these data cannot separate.

We measured the same quantity in 16 screens (Fig. 4c; Supplementary Table S3). Outside the CD4 screen, where constant-region halves showed that the excess agreement of null perturbations comes mostly from clones, and the organoids, where it was measured with clone halves, the quantity is an upper bound on clone strength: the difference between random halves of one null sgRNA and two different null sgRNAs also contains any effect specific to that sgRNA. One screen allows the two to be separated. In the primary T-cell screen of ref.\textsuperscript{25}, the cells come from two donors, so halves from different donors carry the same sgRNAs but share no clone; there the clone part was at most 0.0006, against an upper bound of 0.0013--0.0029. The screens also differ in cell type, type of perturbation and sequencing depth, and most did not record how many cells were transduced per sgRNA, so they can show only whether the orders of magnitude are consistent with the relation above. The upper
bound was highest in cells that passed through an engraftment bottleneck and long expansion after transduction: 0.045--0.10 in human chimeric antigen receptor (CAR) T cells after 21 days in mice\textsuperscript{26} and 0.010--0.015 in mouse T cells 8 days after transfer into infected mice\textsuperscript{27}; in the genome-scale primary CD4 screen, where it is mostly clonal, it was 0.004--0.012, highest in the donor whose perturbations were most clonal by the TRBC marker. It was lowest in non-dividing neurons\textsuperscript{28}: the sgRNAs were delivered by intravenous adeno-associated virus (AAV) into mice 16 days after birth, and the cells sequenced weeks later were neurons that no longer divide, so the transduced cells had no descendants; the upper bound was 0.0001--0.0002. The other systems, including seven cell-line screens\textsuperscript{1,2,4,29}, a primary T-cell screen with only 146 sgRNAs\textsuperscript{25}, T cells transferred into tumour-bearing mice\textsuperscript{30},
primary leukaemia cells\textsuperscript{31}, human brain organoids\textsuperscript{21} and embryonic mouse cortex\textsuperscript{32}, lay between 0.0002 and 0.003 (the organoid and cortex screens have a single null sgRNA and were estimated with clone halves or library halves; Methods). Primary cells are not uniformly high: among primary T cells, the clone part of the small-library screen was at least seven times lower than clone strength in the genome-scale CD4 screen. What sets the size thus appears to be not whether the cells are primary or in vivo but how many effective clones a perturbation's cells come from; the relation was checked in the screen that recorded clones, and the upper bounds of the other screens agree with it in order of magnitude.

Clone strength also has to be read against the size of the real effects. In CD4, the random-half signal of non-targeting sgRNAs was comparable to that of a typical targeting perturbation (0.8--1.1 times); in the CRISPRi screens of cell lines\textsuperscript{1,2} and in the H1 embryonic stem cell data of the 2025 Virtual Cell Challenge\textsuperscript{4} it was only 1--20\% (Supplementary Table S3). The same clone strength therefore matters more in screens whose real effects are weak.

\subsection{Ceilings and reliability labels are inflated}

\begin{figure}[!tbp]
\sbox{\tbox}{\begin{minipage}{\textwidth}\makebox[\textwidth][c]{\includegraphics[width=170mm]{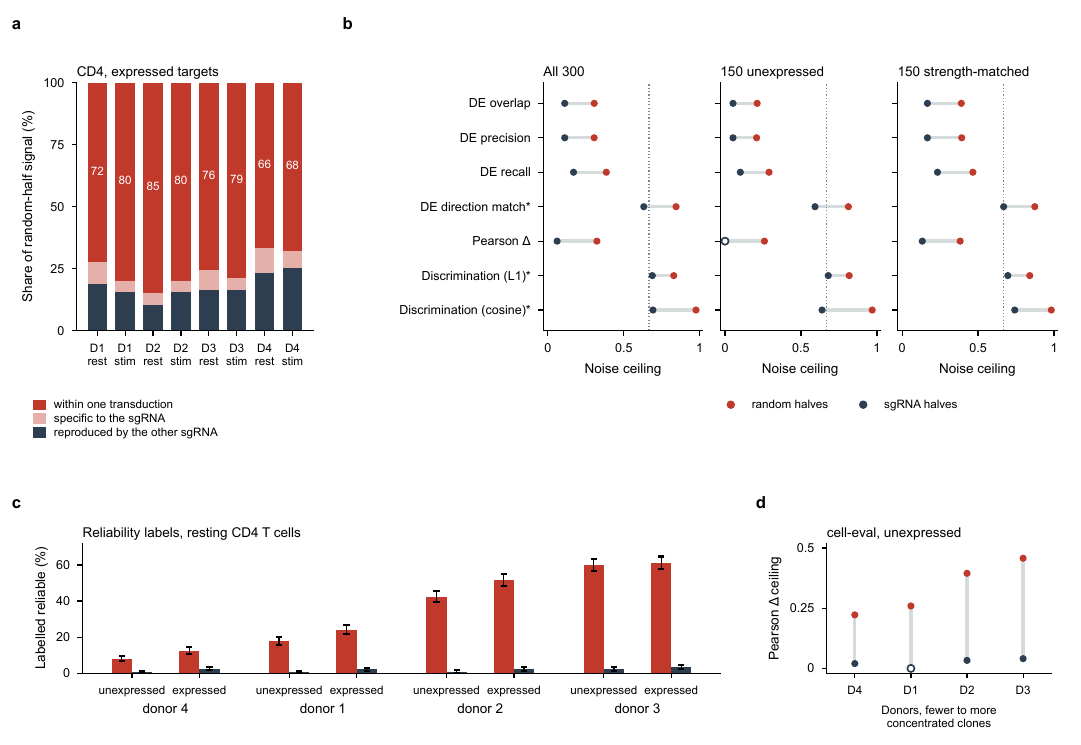}}
\caption{\textbf{Ceilings and reliability labels are inflated.} For the average expressed target, most of the signal that random halves call reproducible is present only between cells of one transduction, which inflates benchmark ceilings and reliability labels based on random halves. \textbf{a}, The random-half cross-half signal of expressed CD4 targets (real effects), set to 100\%, split into three parts. Dark blue, the part reproduced by the other sgRNA: the sgRNA-half cross-half signal as a share of the random-half signal. Light red, the part specific to the sgRNA, estimated from the difference between the agreement of the same sgRNA across two donor groups and that of the two sgRNAs (Methods). Red, the rest, shared only between cells of one transduction. Numbers in the red segments, its percentage. n = 787 resting and 847 stimulated targets (donor 1), 515 and 536 (donor 2), 542 and 539 (donor 3), 1,066 and 1,237 (donor 4). \textbf{b}, A benchmark ceiling is the score a perfect model could
reach on a metric. The evaluation tool cell-eval (version 0.8.2) computes it by splitting each perturbation's cells into random halves, computing the metric with one half as the ``prediction'' and the other as the ``answer'', and correcting from half to full cell number with the Spearman--Brown formula. Red, the ceiling from random halves (mean of three seeds); dark blue, the same computation with the two halves taken from the two sgRNAs (mean of both directions). 300 perturbations of resting CD4+ T cells of donor 1 (left; the other donors in panel d and Supplementary Table S4c), of which 150 unexpressed targets (middle) and 150 expressed targets whose two sgRNAs knocked down the target to a similar extent (right); median 62 cells per half. Metrics as defined by cell-eval (all values in Supplementary Table S4). Asterisks, metrics with chance level 0.5; dotted line, the corrected chance level, 0.67; open point, ceiling not defined (split-half correlation at most 0), plotted at 0. On
every metric the random-half ceiling exceeds the sgRNA ceiling, and unexpressed targets, which should have no effect, receive substantial ceilings under random halves. DE, differential expression.}\label{fig:5}\end{minipage}}
\typeout{FIG 5 \the\dimexpr\ht\tbox+\dp\tbox\relax\space\the\textheight}\usebox{\tbox}
\end{figure}
\begin{figure}[!tbp]
{\small\textbf{Figure 5 (continued).} \textbf{c}, Reliability labels: under a split-half rule, a perturbation is labelled reliable if the correlation between its two half-effects, corrected by the Spearman--Brown formula, is at least 0.5. Percentage of unexpressed and expressed targets labelled reliable under random halves (red) and sgRNA halves (dark blue), resting cells. Donors are ordered from the least to the most concentrated clones by the TRBC readout (Supplementary Table S2b). n = 1,459, 1,235, 869 and 807 unexpressed and 1,421, 1,097, 859 and 745 expressed targets in donors 4, 1, 2 and 3; error bars, bootstrap 95\% intervals over targets. \textbf{d}, The Pearson Δ ceiling of panel b for the unexpressed targets of each donor, in the same order: random halves (red) and sgRNA halves (dark blue; open point, not defined, plotted at 0). The random-half ceiling rises with clone concentration; the sgRNA ceiling does not.\par}
\end{figure}

The shared component appears not only in null perturbations but also in real effects. For expressed targets, whose knockdown has real effects, the random-half signal in CD4 can be split into three parts: 10--25\% is reproduced by the other sgRNA, 4--10\% is specific to the sgRNA and 66--85\% is present only between cells of one transduction, in each of the four donors and both states (Fig. 5a; Methods). A second, independent estimate comes from constant-region halves, which share the sgRNA but not the clones and so keep only the parts related to the sgRNA: among the cells that could be assigned a constant-region class, they kept 26--44\% of the random-half signal, somewhat more than the 15--34\% of the first two parts together in each donor and state, which may come from the few cells assigned to the wrong class letting some clones into both halves. Both estimates show that, for the average expressed target, most of what random halves call reproducible is clonal. Many expressed targets
have weak effects, however. When the targets present in all four donors were divided into thirds by the strength of their effect in the other three donors (measured between sgRNAs, so free of shared clones), the share present only within one transduction was at least 56\% in the weakest third (above 100\% in one case, where the reproduced part was slightly negative) and 5--85\% in the strongest, varying widely between donors and states (about 20 targets per third; Methods).

Where no independent replicate is available, a screen's own null perturbations allow an approximate correction of the average signal. Subtracting the mean random-half signal of the unexpressed targets from that of the expressed targets reduced it from 4--10 times the sgRNA-half signal to 0.5--1.5 times the part related to the sgRNA (the first two parts above), in all four donors and both states (Methods). The correction thus brings the average close to the right size, but only to within about a factor of two; because the clone part differs between perturbations (Fig. 3b), subtracting one constant does not correct the ceiling or the label of an individual perturbation. We next asked what this means for benchmark ceilings and reliability labels.

Benchmarks use random halves to set the score that a perfect model could reach. The evaluation tool cell-eval (version 0.8.2)\textsuperscript{7} computes this ceiling by comparing two random halves of each perturbation on each of its metrics and correcting the result from half to full cell number with the Spearman--Brown formula. We ran the tool on 300 perturbations from resting CD4+ T cells of each donor (299 in donor 2): 150 unexpressed targets and 150 strength-matched knockdowns, in which both sgRNAs knocked down the target to a similar extent, with equal numbers of cells per sgRNA. We compared its random-half ceiling with the same metrics computed between the two sgRNAs of each perturbation.

The random-half ceiling exceeded the sgRNA ceiling on every metric (Fig. 5b; Supplementary Table S4). In donor 1, for the Pearson correlation of effects it was 0.32 against 0.06 over all 300 perturbations, and 0.38 against 0.13 for the strength-matched knockdowns, so the gap is not explained by one sgRNA being weaker. Perturbations with no expected effect also received high ceilings, for example 0.26 for the Pearson correlation and 0.97 for discrimination between perturbations, while their sgRNA ceilings stayed near the corrected chance level. The same held in the other three donors: in all four, the random-half ceiling exceeded the sgRNA ceiling on each of the seven metrics of Fig. 5b and in each subset (84 of 84 comparisons; Supplementary Table S4c). How far it exceeded it followed the clone concentration given by the TRBC marker: for unexpressed targets, the Pearson ceiling was 0.22, 0.26, 0.40 and 0.46 in donors 4, 1, 2 and 3, ordered from the least to the most clonal, while the
sgRNA ceiling stayed at 0.04 or below (Fig. 5d). For metrics whose chance level is 0.5, the Spearman--Brown correction itself turns chance into 0.67 (Supplementary Fig. S2); this is a separate problem, which applying the correction to the score above chance would avoid.

Reliability labels behave the same way. Under a recently proposed split-half rule\textsuperscript{6}, a perturbation is labelled reliable if the corrected correlation between its halves is at least 0.5. With random halves, 8--60\% of the unexpressed and 13--61\% of the expressed targets were labelled reliable in the four donors, against 0.6--2.4\% and 2.2--3.4\% with sgRNA halves (Fig. 5c). The random-half share followed the clone concentration given by the TRBC marker: for unexpressed targets it was 8\% in donor 4, whose perturbations were the least clonal, 18\% in donor 1, 42\% in donor 2 and 60\% in donor 3, the most clonal (Supplementary Table S2a,b). A split-half measure from an independent software package behaves the same way. The split-half reproducibility of Shesha (version 0.2.29)\textsuperscript{33,34}, the cosine between the shift vectors of two random halves relative to the control centroid, had a median of 0.27 for unexpressed targets in resting cells of donor 1 (n =
1,438) and 0.30 for expressed targets (n = 229); computed in the same way between the two sgRNAs of each target, it was 0.002 and 0.05, and 99.6\% of the unexpressed targets scored higher under random halves (Methods).

\subsection{Shared clones create cross-context structure}

\begin{figure}[!tbp]
\sbox{\tbox}{\begin{minipage}{\textwidth}\makebox[\textwidth][c]{\includegraphics[width=170mm]{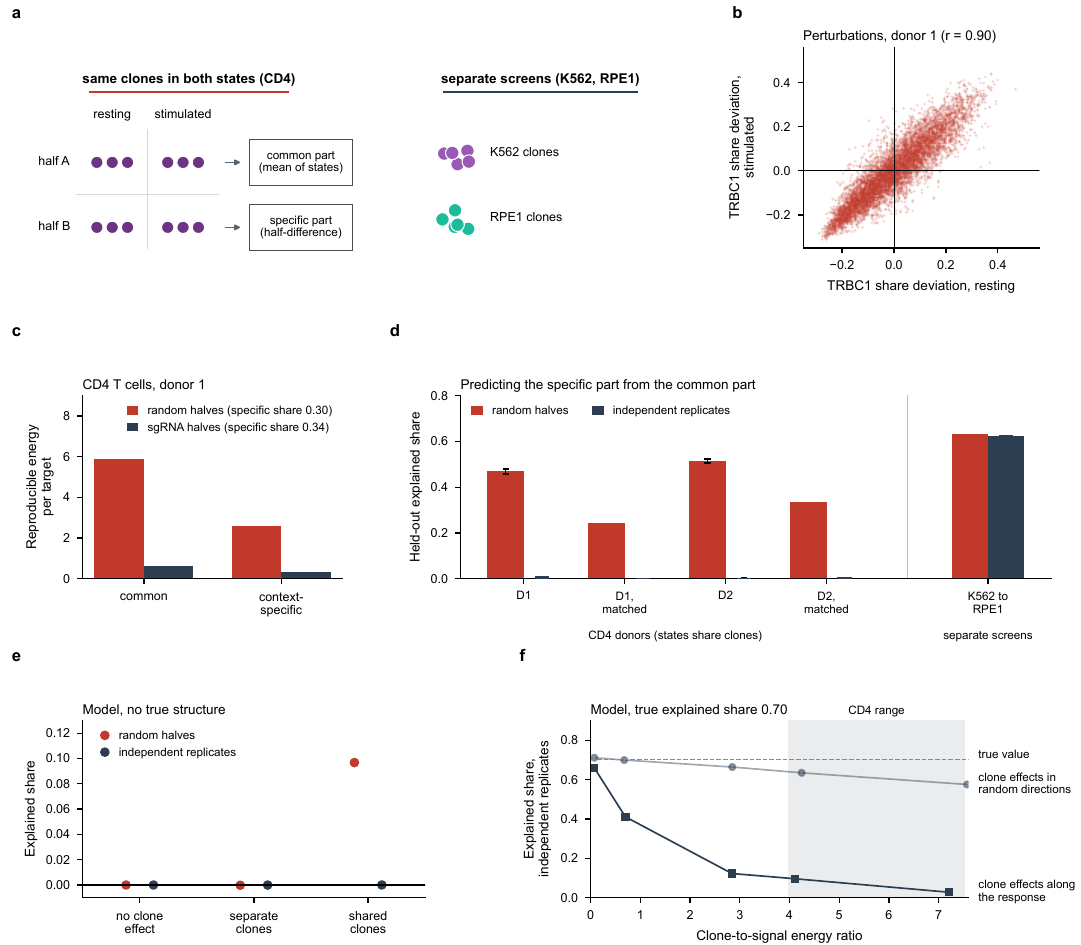}}
\caption{\textbf{When two states share clones, random halves create cross-context structure that independent replicates do not reproduce.} \textbf{a}, Schematic. Left, CD4: a perturbation's cells were expanded and then divided into a resting and a stimulated portion, so both states contain the same clones (purple). The cells of each state were split at random into halves A and B, giving four sets of cells, each containing these clones. The common part is the mean of the effects in the two states, computed from half A; the specific part is half the difference, computed from half B. If a clone shifts expression by c at rest and by qc after stimulation (q ≠ 1), it adds (1 + q)c/2 to the common part and (1 − q)c/2 to the specific part; both halves contain the clone, so the common part of half A and the specific part of half B are correlated through c.~Right, K562 and RPE1: the two cell lines were screened separately, and the cells of the two contexts come from different clones.}\label{fig:6}\end{minipage}}
\typeout{FIG 6 \the\dimexpr\ht\tbox+\dp\tbox\relax\space\the\textheight}\usebox{\tbox}
\end{figure}
\begin{figure}[!tbp]
{\small\textbf{Figure 6 (continued).} \textbf{b}, Each point is a perturbation: horizontal axis, its TRBC1 deviation in resting cells; vertical axis, in stimulated cells. The deviation is the mean TRBC1 share of the perturbation's cells minus the population mean, as in Fig. 3c; at least 30 cells with TRBC counts in each state. If the cells of the two states came from different transductions the deviations would be unrelated, as for the two sgRNAs in Fig. 3c (right; r ≈ 0); the observed r = 0.90 is close to the 0.93 between random halves of the same cells (Fig. 3c, left), so the two states contain the same clones. Donor 1, n = 6,582 perturbations. \textbf{c}, Reproducible energy of the common and the specific parts: the cross-half signal of each part between halves A and B, averaged over targets. Under random halves (red) both parts are 8--9 times larger than under sgRNA halves (dark blue), while the share of the specific part is nearly the same (key: 0.30 and 0.34). Donor 1, n = 6,615 targets. \textbf{d}, Predicting the
specific part from the common part. A linear predictor learns on some perturbations how the common part of half A relates to the specific part of half B and is scored on the others; vertical axis, the share of the reproducible energy of the specific part that it explains on those perturbations. Left four groups, CD4: all targets of two donors, and expressed targets whose two sgRNAs knocked down the target to a similar extent (matched); random halves (red) and sgRNA halves (dark blue). Right group, K562 and RPE1, screened separately: red, random halves within each screen; dark blue, the K562 side taken from another K562 screen (the RPE1 side still comes from random halves of one screen and shares its clones). The two agree, so with the weak clones of these cell lines, clones do not drive this structure. Error bars, bootstrap 95\% intervals over targets (computed for all targets only). \textbf{e}, Simulation. Screens were generated from a simple model: the expression of a cell =
perturbation effect × (1 + its clone's gain) + its clone's state + sampling noise. A clone's state is an expression deviation common to its cells and unrelated to the perturbation; its gain scales its response to the perturbation up or down (Methods). In this simulation the specific part of a perturbation is unrelated to its common part, so the true explained share is 0. Three scenarios: no clone effect; the two contexts with cells from different clones; the two contexts sharing clones. Vertical axis, the share explained by the predictor of panel d on unseen perturbations, under random halves (red) and independent replicates (dark blue), means of three seeds. Only when the contexts share clones do random halves give 0.10; independent replicates give 0 in all three scenarios. \textbf{f}, The same model, now with a real cross-context relation built in: part of each perturbation's specific part is a linear function of its common part, set so that, without clone effects, the predictor
explains 70\% of the specific part's reproducible energy (dashed line, ``true value''). Clone effects of increasing strength were then added, and the explained share was measured between independent replicates (both lines, dark blue; light and dark distinguish only the direction of the clone deviations). Each point is one clone strength (standard deviation of the clone states 0.2, 0.6, 1.2, 1.5 and 2.0; means of three seeds). Horizontal axis, the strength of clones: the null cross-half signal under random halves divided by the cross-half signal of the real effects between independent replicates; grey band, the range of the two CD4 donors (4.0--7.5). When the clone deviations point in random directions (light), only a small part of them falls into the few directions of the perturbation responses among a thousand genes, and the measured value stays near the true value. When they point along the responses (dark), they fall exactly into the directions the predictor uses, swamp the real
relation and leave only 0.10 and 0.03 within the CD4 range. A near-zero value for CD4 sgRNA halves in panel d therefore has two possible explanations: CD4 has no such cross-context relation, or it has one that clone effects hide. These data cannot distinguish the two.\par}
\end{figure}

Perturbation effects measured in several cellular contexts, such as cell types or states, are often split into a part common to all contexts and a part specific to each context\textsuperscript{10}. For two states, resting and stimulated, in which a perturbation has effects x\textsubscript{R} and x\textsubscript{S}, the common part is s = (x\textsubscript{R} + x\textsubscript{S})/2 and the specific part is d = (x\textsubscript{R} − x\textsubscript{S})/2. If the contexts share clones, the deviation of a clone enters both parts (Fig. 6a). (Sharing between contexts is not even necessary: if the random halves share clones within each context and the clone energies differ between contexts, the common part of one half and the specific part of the other are still related, by a quarter of that difference in expectation.) This is the case in the CD4 screen: after transduction the cells were expanded and then divided into two portions, one kept at rest and one restimulated for 48
hours\textsuperscript{3}, so both states contain descendants of the same clones. The TRBC marker confirms this: the TRBC1 deviation of a perturbation correlated between the two states nearly as strongly as between random halves (r = 0.90 in donor 1, Fig. 6b; 0.88--0.96 in the four donors, Supplementary Table S2a). Under both kinds of split, the specific part made up about 0.3 of the reproducible energy, but under random halves the energies of both parts were 8--9 times larger in donor 1 (Fig. 6c). A stable share therefore does not show that an estimate is free of shared-clone effects.

Shared clones make the common part appear to predict the specific part. To predict a perturbation's effect in a context where it has not been measured, two things are available: the perturbation's effects in the measured contexts, and the effects of other perturbations in the new context. Decomposition methods use the first to estimate the perturbation's common part and the second to learn how the common part maps to the part specific to that context. The key step of the transfer is thus to predict the specific part from the common part, and this is the step we tested. We trained a linear predictor f on many perturbations to predict the specific part computed from one half of the cells (B) from the common part computed from the other half (A), d\textsubscript{B} ≈ f(s\textsubscript{A}), and scored it on perturbations it had not seen. Taking the two parts from different halves is meant to keep sampling noise from linking them. Random halves keep the noise apart but not the clones: the
halves contain the same clones, these clones are present in both states, and a clone's deviations in the two states are related but not identical. If a clone shifts expression by c at rest and by qc after stimulation (q ≠ 1), it adds (1 + q)c/2 to s and (1 − q)c/2 to d, and when A and B contain the same clone, f can predict d\textsubscript{B} from s\textsubscript{A} through c alone (Fig. 6a). In CD4, the predictor explained about half of the reproducible specific energy under random halves and almost none under sgRNA halves (Fig. 6d), and it carried the deviations of unexpressed targets into its predictions as much as those of expressed targets (Supplementary Table S5). Between K562 and RPE1, which were screened separately, the predictor explained a similar share with random halves and with the K562 side taken from another screen (Fig. 6d), so the analysis does detect a cross-context relation where clones are weak. In CD4, what it learned was the state of the clones, not the response
to the perturbation. A linear predictor is used here because the leakage caused by clones is itself linear: a clone adds to the specific part exactly (1 − q)/(1 + q) times what it adds to the common part. Its failure across independent replicates shows only that there is no linear relation that reproduces across transductions; published nonlinear models are tested in the next section.

A model with three terms reproduces these observations qualitatively (Fig. 6e,f; Supplementary Fig. S1; Methods). Each cell carries an additive state of its clone, a multiplicative gain of its clone on the response to the perturbation, and sampling noise. The additive term makes null perturbations reproducible and raises the ceilings; neither clone term shrinks with more cells; and when contexts share clones, the predictor explains 10\% of the specific energy even when the true cross-context relation is zero. The model also shows the opposite risk: large clone deviations along the directions of the response can hide a real cross-context structure between independent replicates. At the ratios of clone energy to signal energy found in resting cells of CD4 donors 1 and 2 (4.0 and 7.5; Methods), a true explained share of 70\% is measured as 10\% and 3\% between independent replicates. When the predictor gives almost nothing under sgRNA halves, there are therefore two possibilities: the CD4
screen has no such cross-context structure, or it has one that clone deviations hide between independent replicates. These data cannot distinguish the two.

\subsection{Shared clones raise the scores of transfer models}

\begin{figure}[!tbp]
\sbox{\tbox}{\begin{minipage}{\textwidth}\makebox[\textwidth][c]{\includegraphics[width=170mm]{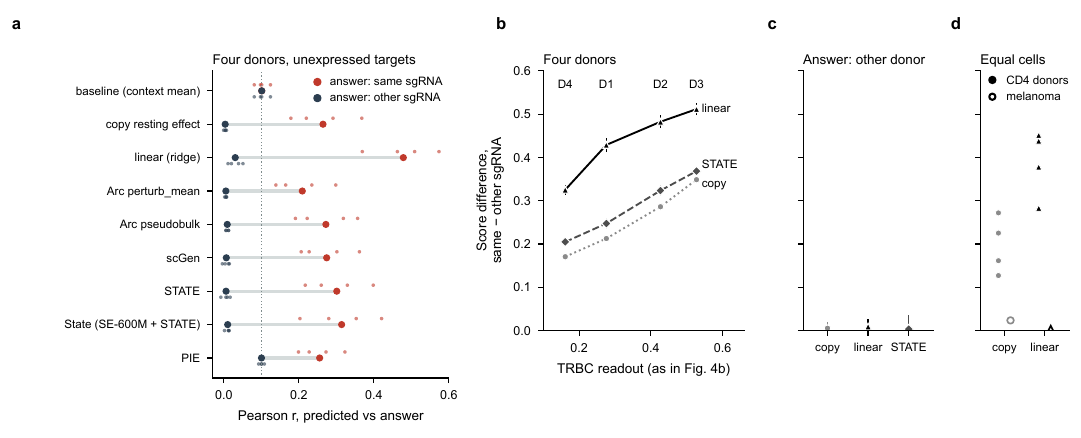}}
\caption{\textbf{Transfer models score higher when the answer shares clones with the input, and the gain disappears when the answer comes from another transduction.} \textbf{a}, Unexpressed targets (no real effect) in the four donors. The training data are the cells of 1,000 targets in both states and the resting cells of the sgRNAs of the test targets; the models predict the stimulated effects of these sgRNAs. The answer is the stimulated cells of the same sgRNA (red; shares clones with the resting cells the model has seen) or of the other sgRNA against the same target (dark blue; another transduction), with at most 200 cells each. Horizontal axis, Pearson correlation over about 2,150 genes between the predicted and the answer effect, averaged over test sgRNAs. Large points, means over the four donors; small points, the individual donors (red above and dark blue below each row). The baseline ignores the perturbation and predicts the mean stimulated effect of the training targets (dotted line,
mean over donors); it is computed against the same control cells as the answers, which raises its score, so it is shown for reference only. Predictors: copying the resting effect, and linear regression with the resting effect as input, are two simple predictors; Arc perturb\_mean (the perturbation's mean effect in the measured contexts) and Arc pseudobulk are models shipped with the State code; scGen\textsuperscript{35} shifts each cell's latent representation by the mean difference between states; STATE is the State transfer model, and State its full pipeline, which first converts cells into embeddings with the foundation model SE-600M and then predicts with STATE; PIE\textsuperscript{36} predicts from evidence about the perturbation in the training contexts and from knowledge sources. Scores of each donor and of knockdowns in Supplementary Table S6a. \textbf{b}, Does the gain grow with the concentration of clones? Each point is a donor (labelled D1--D4 at the top), and the four
donors of one predictor are joined by a line labelled at its end: copying the resting effect (copy, dotted), linear regression (linear, solid) and STATE (dashed). Vertical axis, the difference between the scores with the two answers (same sgRNA minus other sgRNA; scores as in panel a), averaged over all test targets of each donor (250--277, about half unexpressed targets and half knockdowns). Horizontal axis, the donor's TRBC clone readout, as in Fig. 4b (mean of the two states); larger values mean that each perturbation's cells are concentrated in fewer clones. The difference may contain effects specific to the sgRNA as well as clones, but the former should not change with the donor's clone readout. All three differences rise in the order D4 \textless{} D1 \textless{} D2 \textless{} D3, the order of the clone readouts. Error bars, standard errors clustered by target. \textbf{c}, Does the difference come from the sgRNA or from clones? In panel b, the same-sgRNA answer shares both the
sgRNA and the clones with the input. Here the answer is the stimulated cells of the same sgRNA in another donor: the same sgRNA sequence, but another transduction and no shared clone; the other-sgRNA answer is also taken from that donor. A difference due to the sgRNA itself would remain; a difference due to clones would disappear. Vertical axis as in panel b; horizontal axis, the three predictors (marker shapes and shades as in panel b). Points, means over 12 donor pairs (input from one donor, answer from another); vertical lines, ranges. The difference is close to zero (0.004--0.010), so the difference in panel b comes from shared clones.}\label{fig:7}\end{minipage}}
\typeout{FIG 7 \the\dimexpr\ht\tbox+\dp\tbox\relax\space\the\textheight}\usebox{\tbox}
\end{figure}
\begin{figure}[!tbp]
{\small\textbf{Figure 7 (continued).} \textbf{d}, Does the difference disappear when clones are not concentrated? Vertical axis as in panel b; horizontal axis, two predictors. Four kinds of marker: shape and shade give the predictor, as in panel b (grey circles, copying the resting effect; black triangles, linear regression); filled and open give the data (filled, CD4, one point per donor, four in all; open, a screen of a melanoma cell line, whose standard errors are smaller than the markers). The cells of each sgRNA were first subsampled to the same numbers (48 resting and 75 stimulated). The melanoma screen also divided the cells of one transduction among treatments, but about a thousand cells were transduced per sgRNA, so its clones are likely not concentrated (a lower ICC could also contribute). With equal cell numbers the difference remains large in CD4 (copying, 0.13--0.27; linear regression, 0.28--0.45) and is close to zero in the melanoma screen (0.02 and 0.004).\par}
\end{figure}

Published transfer models can also take shared clones for predictive ability. A model sees a perturbation's cells in the resting state and is then scored against the same perturbation's cells in the stimulated state, and in the CD4 screen these two sets of cells come from the same clones.

We tested this in each of the four donors with the evaluation used for the few-shot setting of published transfer models\textsuperscript{11}: some perturbations are held out in the target context and all others are used for training (Methods). The training data were the cells of 1,000 targets in both states, together with the resting cells of the sgRNAs of 250--277 test targets; the models predicted the stimulated effects of these sgRNAs. There were two answers: the stimulated cells of the same sgRNA, which share clones with the resting cells the model has seen, and the stimulated cells of the other sgRNA against the same target, which come from another transduction and share no clone. Each answer used at most 200 cells of its sgRNA. The score was the correlation between the predicted and the answer effects over about 2,150 genes, with the target gene left out. A baseline that ignores the perturbation predicts the mean stimulated effect of the training targets.

Scored against the same sgRNA, every predictor that uses the sgRNA's own resting cells scored well on unexpressed targets in every donor: averaged over the four donors, STATE\textsuperscript{11} scored 0.30, the full State pipeline (which first converts cells into embeddings with its foundation model and then predicts with STATE) 0.31, scGen\textsuperscript{35} 0.27 and PIE\textsuperscript{36} 0.26 (Fig. 7a; Supplementary Table S6a). The simplest predictors did nearly as well or better: copying the resting effect scored 0.26, and linear regression 0.48, higher than any published model, which shows how much of these scores the sgRNA's own resting cells carry. These targets have no real effect, so the difference between these scores and those against the other sgRNA can only come from shared clones or from effects specific to the sgRNA; the cross-donor control below excludes the latter. Scored against the other sgRNA, the same predictors scored 0.004--0.03 on average, and PIE 0.09--0.11;
for comparison, the stimulated effects of the two sgRNAs of a target agree with each other at r = −0.01--0.04 on unexpressed targets and 0.06--0.08 on knockdowns, a reference level for agreement with the other sgRNA. The baseline, which ignores the perturbation, scored 0.08--0.12 against both answers; its effects and the answers are computed against the same observed control cells, whose sampling error therefore enters both and raises its score, so we do not compare models with its level. Stack\textsuperscript{37}, run in donor 1 only, scored 0.04 against the same sgRNA and 0.003 against the other; CPA\textsuperscript{38} and biolord\textsuperscript{39} with their tutorial settings did not learn per-sgRNA effects in these data and could not be judged (Supplementary Table S6b).

If the difference between the two answers comes from shared clones, it should grow with the concentration of clones and disappear when the answer shares no clone with the input. Both hold. Ordered by the clone concentration that the TRBC marker gives for the four donors, the difference for STATE rose from 0.20 to 0.37, and those for copying the resting effect and for linear regression rose in the same order (Fig. 7b). We then changed the answer to the stimulated cells of the same sgRNA in another donor: the same sgRNA, but from another transduction and with no shared clone. Over the 12 donor pairs (input from one donor and answer from another), the difference averaged only 0.004--0.010 (Fig. 7c). The difference therefore comes from shared clones and not from the sgRNA itself.

Sharing clones between contexts is not enough for a large difference; each perturbation's cells must also be concentrated in few clones. A screen of a melanoma cell line\textsuperscript{29} also divided the cells of one transduction among different treatments, but about a thousand cells were transduced per sgRNA, so its clones are likely many (a lower ICC in this cell line could also contribute). For comparability, we subsampled the CD4 cells of each sgRNA to the same numbers. For copying the resting effect, the difference was then about 0.02 in the melanoma screen and 0.13--0.27 in CD4; for linear regression it was 0.004 and 0.28--0.45 (Fig. 7d).

The problem arises only when the source and the target context come from cells of the same transduction, for example different states split from one culture, different time points of the same cells, or different cell types in one organoid. Both designs are in use (Supplementary Table S7, from the methods of 14 public multi-context screens). The contexts come from one transduction in several screens of immune cells and in the melanoma screen above, which PerturBench uses for a transfer task across conditions\textsuperscript{40}; they are separately transduced in the cell-line screens\textsuperscript{1,2} across which State was evaluated\textsuperscript{11}, and then the source and the target context share no clone. Sharing clones is necessary for the gain but not sufficient: in the melanoma screen each sgRNA's cells likely came from many clones (a lower ICC could also contribute), and the gain was small (Fig. 7d). Among the screens whose clone strength we measured or bounded (Fig. 4c),
the CD4 screen is the only one in which the contexts come from one transduction and each perturbation's cells are concentrated in few clones, so the gains measured here are the strongest case among these screens, not a typical one (the CAR T-cell and mouse T-cell screens, whose upper bounds are higher, have no contexts from one transduction in their single-cell data).

Box 1 summarizes how to check whether random halves can be trusted and how to assess the risk.

\begin{center}\fbox{\begin{minipage}{0.94\textwidth}\small\setlength{\parskip}{0.5em}

\textbf{Box 1 \textbar{} Checking whether random halves can be trusted}

\emph{Do the two halves come from the same transduction?} Random halves of the same cells come from one transduction and may share clones; cells of different sgRNAs, different donors or a separate transduction do not. The latter can be used as replicates, and a null control should still be reported.

\emph{If they come from the same transduction} (random halves, or cells of one transduction divided among contexts), reliability labels, ceilings, cross-context structure and the scores of transfer models can all be inflated. How much depends on the effective number of clones per perturbation; by the relation in the Results, the fewer cells transduced per sgRNA, the longer and more uneven the expansion after transduction and the narrower the bottleneck the cells pass through (for example transfer into an animal), the smaller the effective number of clones and the larger the inflation. In the data measured here, the inflation was large in the genome-scale primary T-cell screen, and the upper bound on clone strength was highest in CAR T cells and in T cells expanded in infected mice; in cell lines the upper bound was about 3 times or more lower than clone strength in the CD4 screen, and in the CRISPRi screens of cell lines also small relative to their real effects; it was lowest in
non-dividing neurons in vivo.

\emph{In every case}, include perturbations that should have no effect, such as non-targeting sgRNAs or unexpressed targets, and compare their agreement between random halves with that between two different null perturbations. Read the random-half signal of the null perturbations against that of a typical perturbation of the same screen: the two were about equal in the CD4 screen, and the null signal was 1--20\% of it in the CRISPRi screens of cell lines (Supplementary Table S3). \emph{For transfer models}, score predictions against cells that share no clone with the input, such as the other sgRNA of the gene or the same sgRNA in another donor. In libraries selected on a growth phenotype, unexpressed targets were selected because their sgRNAs affected growth, most likely by acting at the promoter of a neighbouring gene; their effects are real and they cannot serve as null perturbations, so non-targeting sgRNAs should be used instead. If no independent replicate is available,
subtracting the mean random-half signal of the null perturbations from the mean signal of the other perturbations estimates the average signal that is not clonal; in the four CD4 donors this estimate came within about a factor of two of the part related to the sgRNA. Because the clone part differs between perturbations, this does not correct the ceiling or the label of an individual perturbation. This estimate assumes that null and real perturbations have similar effective numbers of clones. Where possible, a clone barcode in the vector allows clone halves that remove this component directly.

\end{minipage}}\end{center}

\section{Discussion}

Random halves are halves within the same clones, and they answer a narrower question than the one they are usually asked. Three questions can be distinguished. First, how precisely the effect has been estimated in this set of cells, for example whether measuring more cells would sharpen the estimate: random halves are the right tool for this question. Second, whether the effect of the same sgRNA would reappear in another experiment: this needs independent transductions of the same sgRNA. Third, how reproducible the effect of knocking down the gene is: this needs different sgRNAs, and the differences between sgRNAs are then part of the variation to be measured. Benchmark ceilings and reliability labels are meant to answer the second or the third question, but use the tool for the first.

The principle has precedents outside single-cell biology: in psychometrics, split-half coefficients computed within one testing occasion omit errors specific to that occasion and overestimate reliability\textsuperscript{41}, and the ENCODE consortium treats pseudoreplicates, two halves of one sample, as a reference that true replicates are not expected to reach rather than as a ceiling\textsuperscript{42,43}. Pseudoreplication has long been known in single-cell differential expression analysis\textsuperscript{15,16}, false positives caused by clonal expansion have been reported in single-cell CRISPR screens at high multiplicity of infection\textsuperscript{17}, it has been pointed out that random halves measure sampling precision\textsuperscript{18}, and the concurrent study above measures how much random splits inflate supervised prediction in lineage-barcoded and patient data, from how many test cells have relatives in training and how close those relatives are in
expression\textsuperscript{19}. Our results add four points. First, in the common screens at low multiplicity of infection, clones carry no label; with halves that differ only in their clones, we show directly that the excess agreement of random halves comes from clones, and we measure its size. Second, this component does not shrink with the number of cells, so more cells, the remedy that sampling models imply, do not remove it. Third, its size follows a simple relation, the expression differences set by clones divided by the effective number of clones; the relation was checked in a screen that recorded clones, and the upper bounds measured in 16 screens are consistent with its orders of magnitude, so the relation says what the size depends on; in a given data set, it is measured most directly with null perturbations. Fourth, shared clones affect not only reproducibility but also cross-context structure and the evaluation of transfer models: when the input and the answer used for
scoring come from cells of the same transduction, much of a model's advantage can come from clones.

Labelled and unlabelled units are handled very differently. In the whole-brain screen of non-dividing neurons, the cells of each sgRNA are spread over 74 mice, and differences between mice are also shared by random halves; but mice are labelled, referencing each cell to control cells of the same mouse removes them, and their measured contribution was small. Clones carry no label, and no reference within one experiment can remove them; only independent transductions, a larger effective number of clones or a clone barcode in the vector can.

We do not question the conclusions of the original studies. They used appropriate replicates in their own analyses; the CD4 screen, for example, treats donors as biological replicates\textsuperscript{3}. The problem arises downstream: when these data are used to compute benchmark ceilings, to label reliable perturbations and to decompose cross-context effects, the two halves are random halves of the same cells; and when transfer models are evaluated, the training cells and the answer may come from the same transduction.

How much this matters depends on clone strength and on its size relative to the real effects. In the CRISPRi screens of cell lines and in the H1 data of the 2025 Virtual Cell Challenge, the shared component of non-targeting sgRNAs was only 1--20\% of the signal of a typical targeting perturbation, and ceilings computed on those data are likely little affected, as far as their null perturbations show. In the genome-scale primary CD4 screen it was comparable to the signal of a typical targeting perturbation, and in two of the three screens of T cells transferred into mice its upper bound was larger. For a new data set, the most direct check is to measure it with the data's own null perturbations, comparing the agreement between random halves with that between two different null perturbations. The same comparison shows where a remedy is needed. An independent replicate within one screen requires at least two sgRNAs per gene in separate vectors, and some of the most widely used data have
none: the CRISPRi screens of K562 and RPE1 express both sgRNAs of a gene from one vector\textsuperscript{1}. In those screens, however, the clone component was small relative to the real effects, and the null check is enough. Where it is large, in primary cells and in cells expanded in animals, a second sgRNA in a separate vector, a second donor or a clone barcode should be part of the design.

Our analysis has limitations, which we group by the conclusion they affect. On the mechanism: constant-region halves distinguish only two classes of clones, and a few cells are assigned to the wrong class; in the organoids, some clone barcodes are shared by several integrations (18\% of the clones with at least 2 cells carried several sgRNAs). Both place clones that should be separated into both halves, which makes the tests conservative: they can underestimate the role of clones, not overestimate it. On the size: the relation was checked in only one screen that recorded clones, and its clone effect is weak; outside the CD4 screen and the organoids, the values of Fig. 4c are upper bounds that also contain effects specific to each null sgRNA; the 16 screens also differ in cell type, type of perturbation (interference, activation, knockout) and sequencing depth, and most did not record the number of cells transduced per sgRNA, so on these data ``the size depends on the effective number
of clones'' is an explanation consistent with the orders of magnitude; several of them have only 4--10 null sgRNAs, and their values indicate orders of magnitude only. On the consequences: sgRNA halves differ in the sgRNA as well as in the clones, so for expressed targets they understate how well the effect of one sgRNA reproduces, which the strength-matched knockdowns limit; ceilings, reliability labels and transfer models were compared in the four donors of one screen, and the models for one pair of states; Stack, CPA and biolord were run in one donor, and several settings differ from the authors' defaults (Supplementary Table S6). Finally, the three-term model shows only that these observations can arise from clones; it was not fitted to the data and does not give their sizes.

Future screens can build independent replicates into their design. Two or more sgRNAs per gene are already standard, and clone barcodes add little cost\textsuperscript{12,21}; separate transductions and donors cost more but give the most direct test. For existing data, a set of null perturbations is the minimum check, and the number of cells transduced per sgRNA and their expansion after transduction indicate the risk, which the null check then measures.

\section{Methods}

\subsection{Data}

\textbf{Primary human CD4+ T cells.} The genome-scale CRISPRi Perturb-seq screen of Zhu et al.\textsuperscript{3} profiles cells of four donors, resting and stimulated (48 h), with a probe-based readout and two sgRNAs per target. We used the single-cell files of all four donors, without cells flagged as low quality or lacking an assigned sgRNA, and the released pseudobulk counts of all four donors. Panels and checks that use donor 1 alone (Fig. 2a, 3b,c, 5b, 6b,c; the Shesha check; Supplementary Table S4a,b) or donors 1 and 2 (Fig. 2c,d, 6d,f; the gene-set checks) say so; the clone tests in each donor are summarized in Supplementary Table S2a.

\textbf{Other screens.} We used the raw single cells of the K562 essential-gene (day 6), K562 genome-wide (day 8, part 1) and RPE1 essential-gene screens of Replogle et al.\textsuperscript{1}, normalized to 10,000 counts over the genes common to the three screens, log1p transformed and corrected additively for the control mean of each GEM group (one 10x Genomics channel); the Jurkat and HepG2 essential-gene screens of Nadig et al.\textsuperscript{2}; the H1 hESC training data of the 2025 Virtual Cell Challenge\textsuperscript{4} (221,273 cells, 150 targets and 31 non-targeting constructs); the brain organoid screen of Li et al.\textsuperscript{21} with its clone barcodes; and eight further screens for clone strength\textsuperscript{25--32}. Systems, uses and accessions are listed in Supplementary Table S8.

\subsection{Null perturbations}

Null perturbations are of two kinds: sgRNAs against targets whose mean expression in control cells is below 0.01 log1p CP10K (unexpressed targets), and non-targeting sgRNAs. Libraries selected on a growth phenotype are not suitable sources of the first kind. The K562 essential-gene library\textsuperscript{1} contains unexpressed targets that were selected because their sgRNAs affected growth, most likely by acting at the promoter of a neighbouring gene\textsuperscript{44}. In two independent K562 screens (the essential-gene and the genome-wide screen), 85\% of the structure of these targets recurred (Supplementary Fig. S3), so their effects are real. In such libraries we used only non-targeting sgRNAs as null perturbations.

\subsection{Effects and gene filters}

The effect of a perturbation in a set of its cells is their mean log1p CP10K expression (the natural logarithm of 1 plus counts per 10,000) minus that of control cells, the non-targeting sgRNAs. Controls were split like the perturbed cells in CD4; in the Replogle data, both halves used all controls, which cancels after centering. The target gene was set to zero in its own effect, except in the ceiling analyses, which keep it as the benchmark tool does. Unless stated otherwise, effects were centered over perturbations within each half and context. Genes were kept if their control mean log1p CP10K was at least 0.1: in CD4 single cells, in both states for analyses that combine states and in the analyzed state otherwise; in the CD4 pseudobulk, in every donor and state; and in the Replogle data, in all three screens. cell-eval used genes with a resting control mean of at least 0.01, together with the target genes. TCR genes (TRAC, TRBC, TRAV, TRBV, TRGC and TRDC) were excluded from the CD4
null, depth, constant-region and clone-strength analyses.

\subsection{Splits and target sets}

In CD4, each cell carries two random bits, r1 and r2 (seed 21). Random halves split on r1; the bootstrap and the reliability labels also used r2 and r1 \(\oplus\) r2. sgRNA halves split on the parity of the sgRNA number in its name, so that each sgRNA falls on the same side in every file, and controls were split likewise by non-targeting sgRNA. Donor halves sum the pseudobulk counts over donors 1--2 or 3--4, both sgRNAs and both sequencing runs (log1p CP10K of the summed counts minus that of the summed controls; at least 40 cells per half and state). In the Replogle data, the cells of each perturbation were permuted (seeds 0, 1 and 2 for the three screens) and cut into equal halves.

Unexpressed targets are probed targets with control mean log1p CP10K below 0.01; expressed targets have at least 0.5. They are an operational null: their knockdown cannot act through the target gene, although an sgRNA can act elsewhere, for example at a neighbouring promoter. Analyses that combine states (Fig. 2a,b, the specific share, the predictor and the reliability labels) apply the threshold in both states; analyses within one state apply it in that state. In CD4, perturbations needed at least 10 cells on each side of both the random and the sgRNA halves in both states for Fig. 2a (1,239 unexpressed targets in donor 1), and at least 10 on each side of all four halvings in both states for Fig. 2b, the predictor, the specific share and the reliability labels (807--1,459 unexpressed targets per donor).

\subsection{Cross-half signal}

The cross-half signal of a perturbation is Σ\textsubscript{g} x\textsubscript{A,g} x\textsubscript{B,g}, the gene-wise product of its two half-effects summed over genes; with independent noise in the two halves, it is an unbiased estimate of the energy of the reproducible effect. For null perturbations, every perturbation was referenced to the null baseline, the mean effect of the unexpressed targets in each half. Referencing to the mean of all perturbations instead (template removal) makes a null perturbation equal to the negative template in both halves; in the K562 genome-wide screen it raised the share of reproducible null targets from 83\% to 89\%.

Non-targeting sgRNAs were treated as perturbations, with the same random bit per cell and effects referenced to the mean over all kept non-targeting sgRNAs in the half (genes with a mean of at least 0.1 over non-targeting cells, TCR genes excluded; at least 10 cells per half). For halves from different sgRNAs, sgRNAs were paired at random and the first half of one was compared with the second half of the other, at the same depth; the share of positive pairs is the mean over 50 pairings. The same analysis was run in the Replogle, Nadig and H1 data, with one pseudo-perturbation per non-targeting construct.

\subsection{Cell number}

To compare cell numbers on the same targets, we re-aggregated the single-cell files of CD4 donors 1 and 2 (resting and stimulated), with the cells, sgRNA parity and the two random bits above, and one further independent random bit per cell (seed 22). The three bits divide the cells of each sgRNA into eight equal parts. Random splits take m parts for one half and m other parts for the other half, so that both halves contain cells of both sgRNAs; sgRNA splits take 2m parts of the first sgRNA for one half and the same 2m parts of the second sgRNA for the other, so that the halves have the same numbers of cells as under random splits (m = 1--4, that is, 1/8 to 1/2 of the cells per half). Each m averages several disjoint choices of parts. Effects are the mean of the half minus the mean of control cells with the same random parts, and the cross-half signal is referenced to the mean effect of the unexpressed targets. Targets needed at least 4 cells on every side of every split, and all cell
numbers used the same targets (donor 1: 1,178 resting and 1,151 stimulated unexpressed targets; donor 2: 709 and 725). The between-half Pearson correlation is the mean over targets of the per-target correlation between half-effects, and the Spearman--Brown value is 2r/(1 + r).

\subsection{TRBC clone marker}

We used cells with at least two TRBC1 plus TRBC2 counts and their TRBC1 share f = TRBC1/(TRBC1 + TRBC2). For a perturbation with n such cells, z = (mean f − f\textsubscript{0})/(sd\textsubscript{f}/√n), where f\textsubscript{0} and sd\textsubscript{f} are the mean and standard deviation over all cells, so that var(z) ≈ 1 under random assignment. var(z) used perturbations with at least 30 cells, and its permutation value is the mean over five shuffles of perturbation labels. The deviation of a set of cells is mean f − f\textsubscript{0}; its reproducibility is its Pearson correlation between two halves (at least 15 cells per half), and between states it was correlated over the perturbations of the main analysis with at least 30 cells carrying TRBC counts in both states (6,582 in donor 1 and 4,963 in donor 2; Fig. 6b) and, for Supplementary Table S2a, over all perturbations with at least 30 such cells in both states (8,689--10,803 per donor).

If the n cells of a perturbation come from K equally sized clones with intraclass correlation ICC of the TRBC1 share, E{[}z\textsuperscript{2}{]} ≈ 1 + ICC(n/K − 1), so K ≤ n/z\textsuperscript{2} for ICC ≤ 1. We report the median of n/z\textsuperscript{2} over resting perturbations with z\textsuperscript{2} \textgreater{} 1 (80--92\% of them), with all sgRNAs of a perturbation pooled. The screen transduced 30 million cells per donor at a multiplicity of infection of about 0.2 with 23,256 sgRNAs\textsuperscript{3}, about 250 transduced cells per sgRNA and 500 per perturbation by design. Each n/z\textsuperscript{2} bounds K only on average, so the median is an estimate rather than a strict bound; the pooled estimate, mean n divided by var(z), gives 4.8, 2.5, 2.4 and 8.7 for resting cells of donors 1--4.

\textbf{Constant-region halves.} Cells in which only TRBC2 was detected, with at least 4 counts, formed one class; cells with a TRBC1 share of at least 0.5 and at least 2 counts formed the other; the remaining cells were not used in this test. The cells of one clone carry one TCRβ rearrangement, so the two classes share almost no clone. The exception is a cell of a TRBC1 clone in which only TRBC2 counts happened to be detected: for a true TRBC1 share of 0.5 and 4 counts this has a probability of about 6\%. Such cells let the halves share a few clones and make the test conservative. For each target, the two classes formed the two halves, and random halves of the same classified cells, with two independent random bits (seed 31), served as the comparison. Each cell was first referenced to the mean of control cells of the same class, which removes expression differences between the two classes of T cells that are unrelated to the perturbation; genes had a control mean of at least 0.1,
without TCR genes. Each class and each random half needed at least 10 cells. For unexpressed and expressed targets we report the share with a positive cross-half signal, the mean cross-half signal and one-sided paired sign and Wilcoxon signed-rank tests of random minus constant-region halves. About 1.0--1.3 million cells were classified per donor and state.

\subsection{Clone barcodes in brain organoids}

The processed single-cell object of Li et al.\textsuperscript{21} (Zenodo 7083558) contains 28,344 cells of the first batch of eight libraries, each with the sgRNA and cell type assigned by the authors. Amplicon sequencing of the clone barcodes came from ArrayExpress E-MTAB-13148 (82 file pairs of the first batch). Read 1 holds a 16-base cell barcode and a 12-base UMI; in read 2, a 16-base insert follows the fixed sequence GGTTCGCCGG, ends in a fixed G and is followed by GCGCGC, and 98\% of reads had this structure. We took the first 15 bases of the insert as the clone barcode, allowing at most one mismatch in the flanking sequence and no N. Files were downloaded in parts and parsed as they arrived, without storing the reads, and accepted only if both reads had the same number of records.

Clones were called as in the original study. Within a cell barcode and UMI, clone barcodes with fewer than 20\% of the reads of the most abundant one were removed; combinations with fewer reads than 1\% of the library's median maximal UMI read count per cell were removed; and a clone barcode one base away from a more abundant one was merged into it. A cell's clone is its clone barcode with the most UMIs, required to have at least 2 UMIs and at least twice the UMIs of the second. Clones were called for 10,967 cells (39\%). Of the clones with at least 2 cells, 82\% carried a single sgRNA, and in clones with several sgRNAs 86\% of the cells carried the most frequent one; these clones with several sgRNAs held 42\% of the cells in clones with at least 2 cells. A clone was defined as one clone barcode within one perturbation, so a barcode shared by different sgRNAs can only merge different clones of one perturbation, which makes clone halves conservative. 36\% of the clones occurred in more
than one library: the transduced cells expanded before they were divided among organoids, so library halves are not fully independent replicates.

Expression was log1p CP10K for genes with a mean of at least 0.1 over all cells. The deviation of a cell is its expression minus the mean of the cells of other perturbations in the same library and cell type (\texttt{celltype\_cl\_coarse2}); a second reference used the library only, which includes differences in cell-type composition. Perturbations needed at least 60 cells with a called clone (32 perturbations). The cross-half signals of random halves and of clone halves (each clone assigned as a whole to a random half) are means over 1,000 splits; with 50 splits, the order of the two for a perturbation with few cells (DEAF1, 64 cells, the second fewest) changed with the draw. In the permutation test, the clone labels were shuffled among the cells of each perturbation, keeping the clone sizes, 200 times, and clone halves were formed from the shuffled labels; the statistic is the mean over perturbations of the random-half minus the clone-half signal, divided by the per-cell variance.

\subsection{Clone strength and the effective number of clones}

Clone strength is the energy of the deviation that the two halves of a null perturbation share, divided by the per-cell variance. In data with several null sgRNAs, the cells of each null sgRNA were split at random (mean over 20 splits), which gives the random-half signal; the first half of one null sgRNA was paired with the second half of another (50 random pairings), which gives the signal between two different null perturbations; their difference was divided by the per-cell variance of the null cells (the sum over genes of the variance around the mean of null cells of the same sample). Different null perturbations share no clone, so the signal between them removes the common shift that the reference introduces. Each effect was referenced to cells of the same sample (10x channel, library, mouse or donor), with two references: the null cells of the sample (excluding the null sgRNA's own cells), or all assigned cells of the sample (excluding the perturbation's own cells). With few null
sgRNAs, the first reference carries the clone states of the other null sgRNAs, which raises the random-half signal and lowers the signal between different null perturbations; in the second, each sgRNA's clones carry little weight. We report the range over both references. sgRNAs needed at least 30 cells.

Data with a single null sgRNA (the GFP control of Jin et al.; Control2 of the organoid screen) cannot be estimated in this way; for these we used, over all perturbations, the random-half signal minus that of halves sharing no clone: clone halves in the organoids, and library halves for Jin et al.~(different libraries are independent transductions of different batches of embryos, but each library pools several embryos, so this is an upper bound on the clonal part). These two screens have open symbols in Fig. 4c.

The effective number of clones is K\textsubscript{eff} = 1/Σ p\textsubscript{k}\textsuperscript{2}, with p\textsubscript{k} the share of clone k among a perturbation's cells with a called clone. The ICC was estimated by one-way random-effects analysis of variance, with between-clone and within-clone mean squares MSB and MSW summed over genes, V\textsubscript{clone} = (MSB − MSW)/n\textsubscript{0}, n\textsubscript{0} = (N − Σ n\textsubscript{k}\textsuperscript{2}/N)/(K − 1) and ICC = V\textsubscript{clone}/(V\textsubscript{clone} + MSW). The slope of clone strength on Σ p\textsubscript{k}\textsuperscript{2} through the origin was compared with the median ICC over perturbations, both with 95\% intervals from 2,000 bootstrap resamples of perturbations. For the out-of-sample prediction, perturbations were split at random into two halves (500 times), and the median ICC of one half times each perturbation's Σ p\textsubscript{k}\textsuperscript{2} in the other half was compared with its
measured clone strength.

The two clone readouts of CD4 are the TRBC readout, the excess variance of each sgRNA's mean TRBC1 share over its sampling variance, divided by the per-cell variance of the share and averaged over sgRNAs, and the clone strength of null perturbations above (geometric mean of the two references). By the relation, each equals its own ICC divided by the same K\textsubscript{eff}, so their ratio does not depend on the donor. The statistical model, the derivation and five propositions with their symbolic checks and recovery in simulated screens are given in Supplementary Note 1.

\subsection{Ceilings and reliability labels}

We ran cell-eval version 0.8.2\textsuperscript{7} in an isolated environment on resting cells of each donor, with at least 40 cells per sgRNA: 150 unexpressed targets drawn at random and the 150 targets with the largest mean knockdown among those whose two sgRNAs both knocked down the target, the weaker reaching a Δlog1p of at most −0.25 and at least 70\% of the stronger. Each perturbation was capped at 400 cells, 15,000 control cells were drawn, and cells were downsampled to equal numbers per sgRNA (median 62 per side); perturbations with fewer than 10 cells per side were dropped. The random-half ceiling is the tool's \texttt{compute\_ceiling} with seeds 0--2. The sgRNA ceiling runs the tool's \texttt{compute} with one sgRNA's cells as ``real'' and the other's as ``prediction'', in both directions, and applies the tool's Spearman--Brown function to the mean over perturbations. Following ref.\textsuperscript{6}, the reliability ρ is the Spearman--Brown value of the median between-half
Pearson correlation, and perturbations with ρ ≥ 0.5 are labelled reliable; for random halves the median was taken over the three halvings r1, r2 and r1 \(\oplus\) r2, and undefined values counted as not reliable.

To split the random-half signal of expressed targets into three parts, the share reproduced by the other sgRNA is the ratio of the sgRNA-half to the random-half cross-half signal, for the same state and targets. To separate the remainder into sgRNA-specific and transduction-level parts, we formed four pseudobulk effects per target and state from all four donors: sgRNA rank 0 or 1 in donors 1--2 or 3--4. The same-sgRNA design pairs each sgRNA across the two donor groups, and the different-sgRNA design pairs sgRNA 0 with sgRNA 1. Their difference, divided by the different-sgRNA signal, estimates sgRNA-specific energy relative to the signal shared across sgRNAs (0.80 resting, 0.48 stimulated); a random half pools the cells of both sgRNAs and keeps only the fraction f = mean over targets of w\textsubscript{0}\textsuperscript{2} + w\textsubscript{1}\textsuperscript{2} of their specific energy, where w\textsubscript{g} is the share of a target's cells carrying sgRNA g (f = 0.55--0.61 in the
four donors and two states), so the sgRNA-specific part is f times this ratio times the single-cell sgRNA-half share, and the rest is the transduction-level part. The single-cell sgRNA-half signal also contains donor-specific responses, so this transfer tends to overestimate the sgRNA-specific part. For the stratified version, the expressed targets of the clone tests present in all four donors (60 resting, 64 stimulated) were divided into thirds by their mean sgRNA-half cross-half signal in the other three donors of the same state, and the three parts were computed within each third (\texttt{code/a100/decomposition\_by\_effect.py}).

\subsection{Shesha split-half reproducibility}

We installed shesha-geometry 0.2.29 from PyPI\textsuperscript{33} and ran its split-half reproducibility function (\texttt{shesha.bio.split\_half\_reproducibility}; 50 random splits, base seed 320, at least 30 cells) on resting cells of CD4 donor 1: all cells of the unexpressed targets, of 400 expressed targets drawn at random (seed 5) and 20,000 non-targeting cells drawn at random as controls, as log1p CP10K expression of the 2,000 genes with the highest control mean among those kept above (TCR genes excluded). For the comparison with independent replicates, we applied the same cosine to fixed halves formed by the two sgRNAs of each target (at least 15 cells each), with the same control centroid. Targets with both values were compared (1,438 unexpressed and 229 expressed).

\subsection{Gene sets and null subtraction}

The gene-set checks repeated the random-half and sgRNA-half cross-half signals of Fig. 2b (resting cells of donors 1 and 2, the targets of the clone tests) with four gene sets: all genes kept above, the 2,000 of them with the highest control mean, those with a control mean of at least 0.5, and those without mitochondrial (MT-) and ribosomal-protein (RPL, RPS) genes; the probe panel contains few of the latter, so the last set removed 11 genes. For the null subtraction, the mean random-half cross-half signal of the unexpressed targets was subtracted from the mean random-half signal of the expressed targets of the same state and donor, and the result was compared with the sgRNA-half signal plus the sgRNA-specific part of Fig. 5a (all four donors, the targets of the clone tests).

\subsection{Common and specific parts, and the predictor}

For two contexts X and Y, the common part is (X + Y)/2 and the specific part is (X − Y)/2, after centering each context over perturbations. The specific share is the cross-half signal of the specific part divided by the sum of the cross-half signals of both parts. The features of a perturbation are its half-A common part projected on the top k right singular vectors of the training perturbations' common parts (k = 20). A ridge regression maps the features to the specific part in half B, with λ chosen from \{10\textsuperscript{−3}, \ldots, 10\textsuperscript{2}\} × tr(Z\textsuperscript{T}Z)/k by five-fold inner cross-validation, and perturbations were split into five outer folds. The explained share is (Σ\(\Vert\)d\textsubscript{B}\(\Vert\)\textsuperscript{2} − Σ\(\Vert\)d\textsubscript{B} − F\(\Vert\)\textsuperscript{2})/Σ d\textsubscript{A} · d\textsubscript{B}, where F is the held-out prediction and the denominator is the reproducible energy of the specific part. A shuffled control,
permuting the features across perturbations, gave values near zero throughout. For K562 and RPE1, perturbations needed at least 20 cells in all three screens, and the independent common part was (other K562 screen + RPE1 half A)/2. Strength-matched knockdowns in CD4 are expressed targets whose two sgRNAs both knocked down the target as above, in both states (615 targets in donor 1 and 470 in donor 2).

\subsection{Model and simulation}

A cell of perturbation p in context c from clone k has expression μ\textsubscript{pc}(1 + g\textsubscript{k}) + a\textsubscript{k} + e, with clone state a\textsubscript{k} \textasciitilde{} N(0, σ\textsubscript{a}\textsuperscript{2}I), clone gain g\textsubscript{k} \textasciitilde{} N(0, σ\textsubscript{g}\textsuperscript{2}) and noise e \textasciitilde{} N(0, σ\textsubscript{e}\textsuperscript{2}I). For random halves, E{[}x\textsubscript{A} · x\textsubscript{B}{]} = \textbar μ\textbar{}\textsuperscript{2}(1 + σ\textsubscript{g}\textsuperscript{2}/K\textsubscript{eff}) + Gσ\textsubscript{a}\textsuperscript{2}/K\textsubscript{eff}; for independent halves it is \textbar μ\textbar{}\textsuperscript{2}. The simulation used G = 1,000 genes, 800 perturbations (20\% null), true responses of rank 10 and log-normal effect sizes. Each perturbation and transduction has K = 6 clones with Dirichlet(0.5) weights, σ\textsubscript{a} = 0.2, σ\textsubscript{g} = 0.6 and σ\textsubscript{e} = 1, and each
half draws 60 cells from the clones. Clones are shared between contexts (additive part scaled by 0.5 in the second) or separate. In the scenario without true structure (Fig. 6e), the effect scale was 0.1, the specific parts were unrelated to the common parts and clone states lay along the common response directions. In the scenario with true structure (Fig. 6f), part of each specific part was a linear function of the common part, set so that the predictor explains 70\% without clone effects, and σ\textsubscript{a} was 0.2, 0.6, 1.2, 1.5 or 2.0, either in random directions or along the common response directions; σ\textsubscript{a} of 1.5 and 2.0 give clone-to-signal energy ratios of 4.1--4.2 and 7.2--7.6, the range in resting cells of CD4 donors 1 and 2. The clone-to-signal energy ratio is the null cross-half signal under random halves divided by the expressed cross-half signal under independent halves. The predictor was a two-fold version of the one above (k = 10), and results are
means over three seeds.

\subsection{Transfer models}

For each CD4 donor we used resting and stimulated (48 h) cells. Test targets were the unexpressed targets and strength-matched knockdowns of the ceiling analysis whose two sgRNAs each had at least 40 cells in both states; 1,000 further targets were drawn at random for training. Each test sgRNA contributed at most 200 cells per state, each training sgRNA at most 80, and 6,000 non-targeting cells were used per state. Genes were the 2,000 with the highest variance of training-target effects among genes with a control mean of at least 0.05 in both states, together with the measured test targets. Perturbations were defined by sgRNA; the resting cells of test sgRNAs entered training and their stimulated cells were held out. This is the few-shot split of the State code (\texttt{{[}fewshot{]}}: specified perturbations of one cell type are held out and everything else is used for training), with the stimulated state as that cell type.

The effect of an sgRNA is the mean expression of its cells minus that of control cells of the same state. The score is the Pearson correlation over genes between the predicted and the answer effect (target gene excluded), averaged over test sgRNAs, with standard errors clustered by target. The answer is the stimulated cells of the same sgRNA (shared clones) or of the other sgRNA against the same target (another transduction); in the cross-donor control, it is the stimulated cells of the same or the other sgRNA in another donor. Versions, settings and deviations from the authors' defaults are given in Supplementary Table S6, with checks that STATE learned per-sgRNA effects and that CPA and biolord did not. The evaluation genes include four T-cell receptor genes (TRBC1, TRBC2, TRAC and TRGC2), which mark clones; leaving them out lowered the same-minus-other difference of the copy, additive and linear predictors by 0.01--0.02 (2--10\%) in each donor and changed their scores against the
other sgRNA by at most 0.006 (\texttt{code/pilot/pilot\_tcr\_check.py}). The published models and the cell-eval ceilings, whose gene set also contains TRBC1, TRBC2 and TRAC, were computed with these genes included. The melanoma screen\textsuperscript{29} was used in its scPerturb version, with cells carrying a single sgRNA, from control to IFN-γ treatment; sgRNAs needed at least 30 cells in both conditions, and test perturbations were cross-fitted in two folds. For the comparison, the CD4 cells of each test sgRNA were subsampled to the cell numbers of the melanoma data (48 resting and 75 stimulated).

\subsection{Statistics}

Confidence intervals are 2.5th--97.5th bootstrap percentiles over 1,000 resamples of perturbations, unless stated otherwise. They describe variation over perturbations within one experiment; agreement across the available independent replicates (four CD4 donors, two for Fig. 2c,d and 6d,f, and two K562 screens) is reported separately. Sign tests and Wilcoxon signed-rank tests are one-sided and paired over targets or perturbations.

\subsection{Software}

Analyses ran in Python 3.11 with NumPy 2.4, SciPy 1.17, h5py 3.16 and PyArrow 20; the organoid object was exported with R and Seurat; cell-eval 0.8.2 ran in its own environment; figures were drawn with Matplotlib 3.10.

\subsection{Data availability}

All data are public; accessions are listed in Supplementary Table S8. The CD4 Perturb-seq data are available at https://genome-scale-tcell-perturb-seq.s3.amazonaws.com/marson2025\_data, the data of Replogle et al.~as figshare+ article 20029387, the data of Nadig et al.~under GEO accession GSE264667 and the H1 data of the 2025 Virtual Cell Challenge through the Arc Virtual Cell Atlas.

\subsection{Code availability}

The analysis code, the result files behind every figure and table, and a record of the reproduction of all reported numbers from the raw data are available at https://github.com/xuanruihuang/random-halves-not-replicates under the MIT licence and are archived at Zenodo\textsuperscript{45}.

\FloatBarrier
\backmatter

\bmhead{Ethics}
This study analysed only publicly released, de-identified data.

\bmhead{Author contributions}
X.H. conceived the study, performed the analyses and wrote the manuscript.

\bmhead{Competing interests}
The author declares no competing interests.

\clearpage
\section{Supplementary Information}
Contents: Supplementary Note 1 (with Table N1); Supplementary Tables S1--S8; Supplementary Figures S1--S3.

\clearpage
\subsection{Supplementary Note 1 \textbar{} Statistical model of clone strength: propositions and checks}

\textbf{Model.} Consider one perturbation measured on G genes. Its cells come from K clones; clone k has n\textsubscript{k} cells, a share p\textsubscript{k} = n\textsubscript{k}/N of all N cells. Cell i belongs to clone k(i), and its expression vector, after subtracting a fixed reference, is x\textsubscript{i} = μ + c\textsubscript{k(i)} + e\textsubscript{i}. Here μ is the perturbation's true effect; c\textsubscript{k} is the deviation of clone k, independent between clones, with mean zero and variance σ\textsubscript{c}\textsuperscript{2} per gene; e\textsubscript{i} is the cell's own variation and sampling noise, independent between cells, with mean zero and variance σ\textsubscript{e}\textsuperscript{2} per gene. The intraclass correlation is ICC = σ\textsubscript{c}\textsuperscript{2}/(σ\textsubscript{c}\textsuperscript{2} + σ\textsubscript{e}\textsuperscript{2}), the share of the differences between single cells that is set by clones. A null perturbation has μ = 0. Clone
deviations and cell noise are assumed independent of each other and of the perturbation, and the reference is estimated from many cells, so that its error can be neglected (the handling of reference error is described in Methods, Clone strength).

\textbf{Proposition 1 (expected cross-half signal).} Let A and B be two sets of cells of a perturbation with no cell in common, in which clone k has shares a\textsubscript{k} and b\textsubscript{k}. Then E{[}\(\bar{x}_A\) · \(\bar{x}_B\){]} = \textbar μ\textbar{}\textsuperscript{2} + Gσ\textsubscript{c}\textsuperscript{2} Σ\textsubscript{k} a\textsubscript{k}b\textsubscript{k}. If A and B share no clone (independent replicates), the second term is zero and only the energy of the true effect, \textbar μ\textbar{}\textsuperscript{2}, remains. \emph{Reason.} \(\bar{x}_A\) = μ + Σ\textsubscript{k} a\textsubscript{k}c\textsubscript{k} + \(\bar{e}_A\). In the expanded inner product only the terms of the same clone, E{[}c\textsubscript{k} · c\textsubscript{k}{]} = Gσ\textsubscript{c}\textsuperscript{2}, are non-zero; A and B have no cell in common, so \(\bar{e}_A\) and \(\bar{e}_B\) are uncorrelated and do not enter. \emph{Check.} With two clones and symbolic shares, the expansion by sympy
equals the right-hand side.

\textbf{Proposition 2 (random halves and clone strength).} When the N cells are split into two halves by a fair coin, E{[}Σ\textsubscript{k} a\textsubscript{k}b\textsubscript{k}{]} = Σ\textsubscript{k} p\textsubscript{k}\textsuperscript{2} − 1/N to first order, a relative error of about K\textsubscript{eff}/N (E{[}n\textsubscript{kA}n\textsubscript{kB}{]} = n\textsubscript{k}\textsuperscript{2}/4 − n\textsubscript{k}/4 for a binomial split of the n\textsubscript{k} cells of clone k). For a null perturbation, the difference between the cross-half signals of random halves and independent replicates, divided by the per-cell variance G(σ\textsubscript{c}\textsuperscript{2} + σ\textsubscript{e}\textsuperscript{2}), is therefore the clone strength, ICC × Σ\textsubscript{k} p\textsubscript{k}\textsuperscript{2} = ICC/K\textsubscript{eff} with K\textsubscript{eff} = 1/Σ\textsubscript{k} p\textsubscript{k}\textsuperscript{2}, and it does not change with the number of cells. \emph{Reason.} The
number of cells of clone k in the first half is binomial, Bin(n\textsubscript{k}, 1/2), and each half has about N/2 cells, so the leading term is (n\textsubscript{k}/2)(n\textsubscript{k}/2)/((N/2)(N/2)) = p\textsubscript{k}\textsuperscript{2}. \emph{Check.} The leading term was verified with sympy. The estimator was tested in simulated screens with known truth: ICC of 0.01, 0.05 and 0.2, 5--200 clones, coefficients of variation of clone size of 0.5--2, 100 or 400 cells per perturbation and 60 null perturbations. Computed as on the data (random halves minus two different null perturbations, divided by the per-cell variance), clone strength correlated with the true ICC × Σ p\textsubscript{k}\textsuperscript{2} at 0.999, with a ratio of 0.98.

\textbf{Proposition 3 (effective number of clones, transduction and expansion).} If N\textsubscript{0} cells were transduced per perturbation and the clone sizes after expansion have mean m and coefficient of variation CV, then to leading order Σ\textsubscript{k} p\textsubscript{k}\textsuperscript{2} = (1 + CV\textsuperscript{2})/N\textsubscript{0}, that is, K\textsubscript{eff} = N\textsubscript{0}/(1 + CV\textsuperscript{2}). \emph{Reason.} Σ\textsubscript{k} p\textsubscript{k}\textsuperscript{2} = Σ\textsubscript{k} n\textsubscript{k}\textsuperscript{2}/(Σ\textsubscript{k} n\textsubscript{k})\textsuperscript{2} ≈ N\textsubscript{0}E{[}n\textsuperscript{2}{]}/(N\textsubscript{0}m)\textsuperscript{2}, and E{[}n\textsuperscript{2}{]} = m\textsuperscript{2}(1 + CV\textsuperscript{2}). \emph{Check.} Verified with sympy. The relation is approximate: it treats the sequenced cells as a proportional reduction of the clone sizes, and when only a small fraction of cells is sequenced, fewer
than N\textsubscript{0} clones are observed.

\textbf{Proposition 4 (limit of the Spearman--Brown extrapolation).} For a null perturbation with a fixed number of clones and σ\textsubscript{c} \textgreater{} 0, the correlation over genes between the two half-means tends to 1 as the number of cells grows, and so does the Spearman--Brown corrected reliability; yet its effect in an independent experiment is zero. The Spearman--Brown formula extrapolates the agreement within the same clones. \emph{Reason.} With more cells the noise terms of both halves vanish, and both halves tend to the same Σ\textsubscript{k} p\textsubscript{k}c\textsubscript{k}. \emph{Check.} In a simulated null perturbation from 4 clones with an ICC of 0.05, the split-half correlation rose from 0.23 to 0.95 and the corrected reliability from 0.38 to 0.98 as the number of cells rose from 50 to 3,200.

\textbf{Proposition 5 (leakage across contexts).} Let two contexts R and S contain the same clones, with clone k deviating by c\textsubscript{k} in R and by qc\textsubscript{k} in S. Let the common part be s = (x\textsubscript{R} + x\textsubscript{S})/2 and the specific part d = (x\textsubscript{R} − x\textsubscript{S})/2, taken from two sets of cells A and B with no cell in common. Then E{[}s\textsubscript{A} · d\textsubscript{B}{]} = Gσ\textsubscript{c}\textsuperscript{2} (1 + q)(1 − q)/4 × Σ\textsubscript{k} a\textsubscript{k}b\textsubscript{k}. When q ≠ 1 and the two sets share clones this is non-zero, so a predictor of d\textsubscript{B} from s\textsubscript{A} can score positively even without any true cross-context relation; when they share no clone it is zero. \emph{Check.} Verified with sympy.

The formal tests reported in the main text are summarized in Supplementary Table N1.

\begin{table}[!htbp]
\renewcommand{\tablename}{Supplementary Table}
\renewcommand{\thetable}{N1}
\sbox{\tbox}{\begin{minipage}{183mm}
\caption{\textbf{Formal tests of the clone mechanism and of the clone-strength relation.}}
\fontsize{8}{9.5}\selectfont\setlength{\tabcolsep}{3pt}\renewcommand{\arraystretch}{1.1}
\begin{tabularx}{\linewidth}{@{}>{\raggedright\arraybackslash}X>{\raggedright\arraybackslash}X>{\raggedright\arraybackslash}X>{\raggedright\arraybackslash}X@{}}
\toprule
Test & Statistic & Null distribution & Result \\
\midrule
CD4, constant-region halves & random minus constant-region halves, per target & paired sign test; Wilcoxon signed-rank test & sign test p \textless{} 10\textsuperscript{−180}, Wilcoxon p \textless{} 10\textsuperscript{−120} in each donor and state \\
Organoids, clone halves & random minus clone halves, per perturbation & sign test; clone labels shuffled within each perturbation (200 times) & 32 of 32, p = 2 × 10\textsuperscript{−10} (1,000 splits per perturbation); shuffled clone halves 4.98, random halves 4.97, true clone halves 3.43; permutation p = 0.005 \\
Relation: slope equals ICC & slope of clone strength on Σ p\textsubscript{k}\textsuperscript{2} & bootstrap over perturbations (2,000) & slope 0.0096 (0.0085--0.0109); median ICC 0.0104 (0.0034--0.0171); interval of their ratio 0.57--2.70 includes 1; the regression does not remove the −1/N term of Proposition 2, which lowers the slope by a few per cent \\
Relation: out-of-sample prediction & ICC of one half of the perturbations predicts the clone strength of the other & random splits of perturbations (500) & median Spearman 0.80, Pearson 0.83; median measured/predicted 0.94 (quartiles 0.71--1.39) \\
\bottomrule
\end{tabularx}
\end{minipage}}
\typeout{TBL N1 \the\wd\tbox\space\the\dimexpr\ht\tbox+\dp\tbox\relax\space\the\textheight}
\makebox[\textwidth][c]{\usebox{\tbox}}
\end{table}
\clearpage
\begin{table}[!htbp]
\renewcommand{\tablename}{Supplementary Table}
\renewcommand{\thetable}{S1a}
\sbox{\tbox}{\begin{minipage}{183mm}
\caption{\textbf{Non-targeting sgRNAs as null perturbations. a, CD4.} Each non-targeting sgRNA of the CD4 screen is a separate transduction with no target. Effects are referenced to the mean over the kept non-targeting sgRNAs in each half; genes with a mean of at least 0.1 over non-targeting cells, TCR genes excluded; sgRNAs with at least 10 cells per half. Different sgRNAs: the first half of one sgRNA against the second half of another, at the same depth, averaged over 50 random pairings (range over pairings in parentheses). Random halves: bootstrap 95\% interval over sgRNAs. TRBC: variance of the z-score of each sgRNA's TRBC1 share (sgRNAs with at least 30 cells carrying TRBC counts; in parentheses after permuting sgRNA labels) and the Pearson correlation of the TRBC1 deviation between random halves (at least 15 cells per half).}
\fontsize{8}{9.5}\selectfont\setlength{\tabcolsep}{3pt}\renewcommand{\arraystretch}{1.1}
\begin{tabular*}{\linewidth}{@{\extracolsep{\fill}}lrrllllr@{}}
\toprule
Donor, state & sgRNAs & \begin{tabular}[b]{@{}l@{}}Cells per \\ half \\ (median)\end{tabular} & \begin{tabular}[b]{@{}l@{}}Positive, random \\ halves\end{tabular} & \begin{tabular}[b]{@{}l@{}}Positive, \\ different sgRNAs\end{tabular} & \begin{tabular}[b]{@{}l@{}}Mean signal, \\ random / \\ different\end{tabular} & \begin{tabular}[b]{@{}l@{}}TRBC var(z) \\ (permuted)\end{tabular} & \begin{tabular}[b]{@{}l@{}}TRBC \\ deviation, \\ between \\ random \\ halves\end{tabular} \\
\midrule
Donor 1, resting & 739 & 39 & 98.2\% (97.2--99.1) & 49.8\% (42.8--54.7) & 13.9 / −0.04 & 34.4 (0.99) & 0.95 \\
Donor 1, stimulated & 785 & 36 & 96.6\% (95.3--97.8) & 49.8\% (45.2--55.1) & 13.5 / −0.01 & 29.1 (1.01) & 0.94 \\
Donor 2, resting & 578 & 41 & 98.8\% (97.9--99.5) & 48.8\% (43.6--56.7) & 20.2 / −0.13 & 79.9 (0.98) & 0.97 \\
Donor 2, stimulated & 622 & 40 & 98.7\% (97.7--99.5) & 49.8\% (43.4--56.6) & 21.3 / 0.01 & 76.0 (1.00) & 0.97 \\
\bottomrule
\end{tabular*}
\end{minipage}}
\typeout{TBL S1a \the\wd\tbox\space\the\dimexpr\ht\tbox+\dp\tbox\relax\space\the\textheight}
\makebox[\textwidth][c]{\usebox{\tbox}}
\end{table}
\clearpage
\begin{table}[!htbp]
\renewcommand{\tablename}{Supplementary Table}
\renewcommand{\thetable}{S1b}
\sbox{\tbox}{\begin{minipage}{183mm}
\caption{\textbf{b, Cell-line screens.} Raw single cells of the K562 essential-gene, K562 genome-wide (part 1) and RPE1 screens of Replogle et al., the Jurkat and HepG2 screens of Nadig et al.~and the H1 hESC training data of the 2025 Virtual Cell Challenge; one pseudo-perturbation per non-targeting construct.}
\fontsize{8}{9.5}\selectfont\setlength{\tabcolsep}{3pt}\renewcommand{\arraystretch}{1.1}
\begin{tabularx}{\linewidth}{@{}>{\raggedright\arraybackslash}Xrrll@{}}
\toprule
Screen & Constructs & \begin{tabular}[b]{@{}l@{}}Cells per \\ half \\ (median)\end{tabular} & \begin{tabular}[b]{@{}l@{}}Positive, random \\ halves\end{tabular} & \begin{tabular}[b]{@{}l@{}}Positive, \\ different \\ constructs\end{tabular} \\
\midrule
CD4 donor 1, resting & 739 & -- & see a & see a \\
CD4 donor 2, resting & 578 & -- & see a & see a \\
K562 essential (day 6) & 97 & 52 & 74.2\% (64.9--82.5) & 49.1\% (33.3--72.9) \\
K562 genome-wide (day 8, part 1) & 507 & 68 & 71.4\% (67.6--75.0) & 49.9\% (42.7--56.9) \\
RPE1 essential & 113 & 49 & 76.1\% (68.1--84.1) & 50.9\% (37.5--69.6) \\
Jurkat essential & 128 & 44 & 73.4\% (65.6--81.2) & 51.4\% (34.4--64.1) \\
HepG2 essential & 112 & 19 & 59.8\% (50.0--68.8) & 50.0\% (33.9--71.4) \\
H1 hESC, VCC 2025 training data & 31 & 574 & 71.0\% (54.8--87.1) & 49.2\% (13.3--80.0) \\
\bottomrule
\end{tabularx}
\end{minipage}}
\typeout{TBL S1b \the\wd\tbox\space\the\dimexpr\ht\tbox+\dp\tbox\relax\space\the\textheight}
\makebox[\textwidth][c]{\usebox{\tbox}}
\end{table}
\clearpage
\begin{table}[!htbp]
\renewcommand{\tablename}{Supplementary Table}
\renewcommand{\thetable}{S2a}
\sbox{\tbox}{\begin{minipage}{183mm}
\caption{\textbf{Clone tests in each CD4 donor.} Each column is one donor and state; --, computed only for resting cells or once per donor. TRBC clone marker: var(z), the variance over perturbations of the z-scored TRBC1 share (1 under random assignment of cells), over all perturbations with at least 30 cells with TRBC counts (Fig. 3b shows the perturbations of the main analysis, for which it is 36 in resting cells of donor 1); Pearson r of the TRBC1 deviation between the random halves and between the two sgRNAs of a perturbation (Fig. 3c), and between its resting and stimulated cells (all perturbations with at least 30 cells with TRBC counts in both states; Fig. 6b); an estimate of the upper bound on the effective number of clones (median of per-perturbation bounds; Methods). Unexpressed targets: share with a positive cross-half signal under random and sgRNA halves (targets with at least 10 cells on each side of every halving in both states; Fig. 2b). Constant-region halves: the test of Fig.
3d, with random halves of the same classified cells as the comparison. Expressed targets: the three parts of the random-half signal (Fig. 5a). Reliability labels under the split-half rule, resting cells (Fig. 5c).}
\fontsize{7}{8.2}\selectfont\setlength{\tabcolsep}{3pt}\renewcommand{\arraystretch}{1.1}
\begin{tabularx}{\linewidth}{@{}>{\raggedright\arraybackslash}Xrrrrrrrr@{}}
\toprule
Quantity & Donor 1, rest & \begin{tabular}[b]{@{}l@{}}Donor 1, \\ stim\end{tabular} & \begin{tabular}[b]{@{}l@{}}Donor 2, \\ rest\end{tabular} & \begin{tabular}[b]{@{}l@{}}Donor 2, \\ stim\end{tabular} & \begin{tabular}[b]{@{}l@{}}Donor 3, \\ rest\end{tabular} & \begin{tabular}[b]{@{}l@{}}Donor 3, \\ stim\end{tabular} & Donor 4, rest & \begin{tabular}[b]{@{}l@{}}Donor 4, \\ stim\end{tabular} \\
\midrule
\textbf{TRBC clone marker} & ~ & ~ & ~ & ~ & ~ & ~ & ~ & ~ \\
var(z) of the TRBC1 share (1 if no clones) & 34 & 28 & 81 & 78 & 81 & 77 & 17 & 19 \\
r of the TRBC1 deviation, random halves & 0.93 & 0.92 & 0.96 & 0.96 & 0.97 & 0.97 & 0.86 & 0.87 \\
r of the TRBC1 deviation, sgRNA halves & 0.00 & −0.01 & 0.02 & 0.02 & −0.01 & −0.02 & 0.00 & −0.01 \\
r of the TRBC1 deviation, resting vs stimulated & 0.89 & -- & 0.94 & -- & 0.96 & -- & 0.88 & -- \\
Estimated upper bound on K\textsubscript{eff} (median n/z\textsuperscript{2}) & 6.5 & 6.8 & 3.9 & 4.0 & 3.4 & 3.7 & 12.4 & 11.6 \\
\textbf{Unexpressed targets, positive cross-half signal (\%)} & ~ & ~ & ~ & ~ & ~ & ~ & ~ & ~ \\
Random halves & 99 & 97 & 99 & 99 & 100 & 100 & 96 & 97 \\
sgRNA halves & 48 & 54 & 53 & 50 & 55 & 54 & 50 & 50 \\
Targets (n) & 1,235 & 1,235 & 869 & 869 & 807 & 807 & 1,459 & 1,459 \\
\textbf{Constant-region halves, unexpressed targets} & ~ & ~ & ~ & ~ & ~ & ~ & ~ & ~ \\
Positive, random halves (\%) & 98 & 95 & 99 & 98 & 99 & 100 & 94 & 96 \\
Positive, constant-region halves (\%) & 56 & 54 & 61 & 58 & 57 & 61 & 58 & 57 \\
Cross-half signal kept (\%) & 13 & 8 & 14 & 16 & 18 & 18 & 18 & 14 \\
Targets higher under random halves (\%) & 95 & 92 & 96 & 94 & 98 & 97 & 89 & 91 \\
Targets (n) & 1,581 & 1,502 & 968 & 981 & 835 & 768 & 1,889 & 1,673 \\
Expressed targets: signal kept (\%) & 31 & 31 & 28 & 29 & 26 & 32 & 38 & 44 \\
\textbf{Expressed targets, random-half signal (\%)} & ~ & ~ & ~ & ~ & ~ & ~ & ~ & ~ \\
Reproduced by the other sgRNA & 19 & 16 & 10 & 16 & 16 & 16 & 23 & 25 \\
Specific to the sgRNA & 9 & 4 & 5 & 5 & 8 & 5 & 10 & 7 \\
Only within one transduction & 72 & 80 & 85 & 80 & 76 & 79 & 66 & 68 \\
Targets (n) & 787 & 847 & 515 & 536 & 542 & 539 & 1,066 & 1,237 \\
\textbf{Labelled reliable, resting cells (\%)} & ~ & ~ & ~ & ~ & ~ & ~ & ~ & ~ \\
Unexpressed, random halves & 18 & -- & 42 & -- & 60 & -- & 8.0 & -- \\
Unexpressed, sgRNA halves & 0.6 & -- & 0.9 & -- & 2.4 & -- & 0.6 & -- \\
Expressed, random halves & 24 & -- & 52 & -- & 61 & -- & 13 & -- \\
Expressed, sgRNA halves & 2.2 & -- & 2.3 & -- & 3.4 & -- & 2.4 & -- \\
Targets, unexpressed / expressed (n) & 1,235 / 1,097 & -- & 869 / 859 & -- & 807 / 745 & -- & 1,459 / 1,421 & -- \\
\bottomrule
\end{tabularx}
\end{minipage}}
\typeout{TBL S2a \the\wd\tbox\space\the\dimexpr\ht\tbox+\dp\tbox\relax\space\the\textheight}
\makebox[\textwidth][c]{\usebox{\tbox}}
\end{table}
\clearpage
\begin{table}[!htbp]
\renewcommand{\tablename}{Supplementary Table}
\renewcommand{\thetable}{S2b}
\sbox{\tbox}{\begin{minipage}{183mm}
\caption{\textbf{Two clone readouts in the CD4 screen.} TRBC readout: the variance of each sgRNA's mean TRBC1 share beyond its sampling variance, divided by the per-cell variance of the share and averaged over sgRNAs. Expression readout: the clone strength of non-targeting sgRNAs (random-half minus different-sgRNA cross-half signal, divided by the per-cell variance), with effects referenced to the null cells or to all assigned cells of the same sample, and their geometric mean. Both readouts are expected to equal their own ICC divided by the same effective number of clones, so their ratio should not depend on the donor. Donors are ordered by the TRBC readout.}
\fontsize{8}{9.5}\selectfont\setlength{\tabcolsep}{3pt}\renewcommand{\arraystretch}{1.1}
\begin{tabular*}{\linewidth}{@{\extracolsep{\fill}}lrrrrrr@{}}
\toprule
Donor, state & \begin{tabular}[b]{@{}l@{}}TRBC \\ readout\end{tabular} & \begin{tabular}[b]{@{}l@{}}Clone \\ strength, \\ null \\ reference\end{tabular} & \begin{tabular}[b]{@{}l@{}}Clone \\ strength, \\ all-cell \\ reference\end{tabular} & \begin{tabular}[b]{@{}l@{}}Clone \\ strength, \\ geometric \\ mean\end{tabular} & Ratio & \begin{tabular}[b]{@{}l@{}}Null \\ sgRNAs\end{tabular} \\
\midrule
Donor 4, resting & 0.156 & 0.0044 & 0.0044 & 0.0044 & 0.028 & 821 \\
Donor 4, stimulated & 0.166 & 0.0049 & 0.0048 & 0.0049 & 0.029 & 875 \\
Donor 1, resting & 0.277 & 0.0071 & 0.0071 & 0.0071 & 0.026 & 688 \\
Donor 1, stimulated & 0.275 & 0.0080 & 0.0080 & 0.0080 & 0.029 & 723 \\
Donor 2, resting & 0.429 & 0.0103 & 0.0103 & 0.0103 & 0.024 & 541 \\
Donor 2, stimulated & 0.425 & 0.0105 & 0.0105 & 0.0105 & 0.025 & 587 \\
Donor 3, resting & 0.535 & 0.0120 & 0.0120 & 0.0120 & 0.022 & 507 \\
Donor 3, stimulated & 0.519 & 0.0107 & 0.0107 & 0.0107 & 0.021 & 515 \\
\bottomrule
\end{tabular*}
\end{minipage}}
\typeout{TBL S2b \the\wd\tbox\space\the\dimexpr\ht\tbox+\dp\tbox\relax\space\the\textheight}
\makebox[\textwidth][c]{\usebox{\tbox}}
\end{table}
The TRBC readout differs 3.4-fold and the expression readout 2.7-fold between donors, while the ratio varies between 0.021 and 0.029 and falls slightly as the readouts rise.

\clearpage
\begin{table}[!htbp]
\renewcommand{\tablename}{Supplementary Table}
\renewcommand{\thetable}{S3}
\sbox{\tbox}{\begin{minipage}{183mm}
\caption{\textbf{Clone strength in 16 screens.} Outside the CD4 screen and the organoids the values are upper bounds: the estimator also contains any effect specific to each null sgRNA. In the screen of Schmidt et al., whose cells come from two donors, halves from different donors share no clone; the clone part estimated with them was at most 0.0006 (−0.00002 to 0.0006 over the references and states). One estimator for all screens with several null sgRNAs: the random-half cross-half signal of null sgRNAs (mean of 20 splits) minus the cross-half signal between two different null sgRNAs (50 pairings), divided by the per-cell variance of the null cells; at least 30 cells per sgRNA. Ranges are over the two references (null cells or all assigned cells of the same sample) and over states, treatments or donors. Null / targeting: the random-half signal of null sgRNAs relative to that of a typical targeting perturbation of the same screen. Screens with a single null sgRNA were estimated with halves that
share no clone (Methods). Screens are ordered as in Fig. 4c.}
\fontsize{8}{9.5}\selectfont\setlength{\tabcolsep}{3pt}\renewcommand{\arraystretch}{1.1}
\begin{tabularx}{\linewidth}{@{}l>{\raggedright\arraybackslash}Xl>{\raggedright\arraybackslash}Xl@{}}
\toprule
Screen & System & \begin{tabular}[b]{@{}l@{}}Null \\ sgRNAs\end{tabular} & Clone strength & \begin{tabular}[b]{@{}l@{}}Null / \\ targeting\end{tabular} \\
\midrule
Knudsen et al. & human CAR-T cells, 21 days in mice & 4 & 0.045--0.10 & 1.0--3.3 \\
Fagerberg et al. & mouse P14 CD8 T cells, 8 days in LCMV-infected mice & 5 & 0.010--0.015 & 2.0--2.4 \\
Zhu et al. & primary human CD4 T cells, genome-wide CRISPRi, 4 donors, resting and stimulated & 507--875 & 0.0044--0.012 & 0.83--1.05 \\
Schmidt et al. & primary human T cells, CRISPRa, resting and restimulated & 8 & 0.0013--0.0029 & 0.10--0.29 \\
Zhou et al. & mouse OT-I CD8 T cells, 7 days in tumour-bearing mice & 37 & 0.0013--0.0017 & 0.44--0.48 \\
Cao et al. & primary human AML cells, in vitro & 10 & 0.0008--0.0018 & 0.07--0.12 \\
Jin et al. & embryonic mouse cortex, in utero lentivirus & 1 & 0.001--0.003 (library halves, upper bound) & -- \\
Li et al. & human brain organoids, clone barcodes & 1 & 0.0009 (clone halves; 0.0046 with cell-type composition) & -- \\
Nadig et al. & HepG2, essential genes, CRISPRi & 87 & 0.0011--0.0015 & 0.04--0.20 \\
Nadig et al. & Jurkat, essential genes, CRISPRi & 127 & 0.0010--0.0012 & 0.06--0.16 \\
Replogle et al. & RPE1, essential genes, CRISPRi & 112 & 0.0011--0.0012 & 0.02--0.18 \\
Frangieh et al. & melanoma cell line, CRISPR knockout, three treatments & 71--74 & 0.0003--0.0009 & 0.35--0.82 \\
Replogle et al. & K562, essential genes, CRISPRi & 95 & 0.0004--0.0006 & 0.03--0.16 \\
Replogle et al. & K562, genome-wide, CRISPRi & 503 & 0.0003 & 0.09 \\
VCC 2025 & H1 hESC, CRISPRi, public training data & 31 & 0.0002--0.0003 & 0.01--0.02 \\
Shi et al. & mouse whole brain, AAV at postnatal day 16, non-dividing neurons; reference within mouse & 399 & 0.0001--0.0002 & 0.35--0.56 \\
\bottomrule
\end{tabularx}
\end{minipage}}
\typeout{TBL S3 \the\wd\tbox\space\the\dimexpr\ht\tbox+\dp\tbox\relax\space\the\textheight}
\makebox[\textwidth][c]{\usebox{\tbox}}
\end{table}
\clearpage
\begin{table}[!htbp]
\renewcommand{\tablename}{Supplementary Table}
\renewcommand{\thetable}{S4a}
\sbox{\tbox}{\begin{minipage}{183mm}
\caption{\textbf{cell-eval 0.8.2 ceilings for all metrics. a, Spearman--Brown corrected (``ceiling'').} Resting CD4+ T cells of donor 1: all 300 perturbations, the 150 unexpressed targets and the 150 strength-matched knockdowns, with equal numbers of cells per sgRNA (median 62 per side). Random: the tool's compute\_ceiling (disjoint random halves of each perturbation and of the controls), range over seeds 0--2. sgRNA: the tool's compute with one sgRNA's cells as ``real'' and the other's as ``prediction'', range over the two directions. Asterisks, metrics whose chance level is 0.5, which the correction maps to 0.67. Daggers, metrics that cell-eval computes over the whole data set, so they are identical across subsets. Dashes, metrics not returned for a subset. n.d., not defined: the split-half correlation was at most 0; the half-depth correlation is given in parentheses.}
\fontsize{8}{9.5}\selectfont\setlength{\tabcolsep}{3pt}\renewcommand{\arraystretch}{1.1}
\begin{tabularx}{\linewidth}{@{}>{\raggedright\arraybackslash}Xllllll@{}}
\toprule
Metric & \begin{tabular}[b]{@{}l@{}}All: \\ random\end{tabular} & All: sgRNA & \begin{tabular}[b]{@{}l@{}}Unexpressed: \\ random\end{tabular} & \begin{tabular}[b]{@{}l@{}}Unexpressed: \\ sgRNA\end{tabular} & \begin{tabular}[b]{@{}l@{}}Knockdowns: \\ random\end{tabular} & \begin{tabular}[b]{@{}l@{}}Knockdowns: \\ sgRNA\end{tabular} \\
\midrule
DE overlap (top N) & 0.30--0.31 & 0.11 & 0.20--0.22 & 0.05 & 0.37--0.40 & 0.16--0.17 \\
DE overlap (top 50) & 0.30--0.32 & 0.11 & 0.20--0.22 & 0.05 & 0.38--0.41 & 0.17 \\
DE overlap (top 100) & 0.30--0.31 & 0.11 & 0.20--0.22 & 0.05 & 0.38--0.40 & 0.17 \\
DE overlap (top 200) & 0.30--0.31 & 0.11 & 0.20--0.22 & 0.05 & 0.37--0.40 & 0.16--0.17 \\
DE overlap (top 500) & 0.30--0.31 & 0.11 & 0.20--0.22 & 0.05 & 0.37--0.40 & 0.16--0.17 \\
DE precision (top N) & 0.30--0.31 & 0.11 & 0.19--0.22 & 0.05 & 0.39 & 0.16--0.17 \\
DE precision (top 50) & 0.30--0.31 & 0.11 & 0.20--0.22 & 0.05 & 0.39--0.40 & 0.17 \\
DE precision (top 100) & 0.30--0.31 & 0.11 & 0.19--0.22 & 0.05 & 0.39 & 0.17 \\
DE precision (top 200) & 0.30--0.31 & 0.11 & 0.19--0.22 & 0.05 & 0.39 & 0.16--0.17 \\
DE precision (top 500) & 0.30--0.31 & 0.11 & 0.19--0.22 & 0.05 & 0.39 & 0.16--0.17 \\
DE recall & 0.36--0.40 & 0.15--0.18 & 0.28--0.30 & 0.09--0.11 & 0.43--0.49 & 0.21--0.26 \\
DE direction match* & 0.84--0.85 & 0.62--0.64 & 0.81--0.82 & 0.57--0.61 & 0.86--0.88 & 0.67 \\
DE Spearman (significant genes)† & 0.80--0.82 & 0.46--0.47 & 0.80--0.82 & 0.46--0.47 & 0.80--0.82 & 0.46--0.47 \\
DE Spearman of log fold changes† & -- & -- & -- & -- & 0.42 & n.d. (r = −0.05) \\
PR-AUC & -- & -- & -- & -- & 0.39--0.43 & 0.13--0.16 \\
ROC-AUC* & -- & -- & -- & -- & 0.84--0.86 & 0.76--0.77 \\
Pearson Δ & 0.32 & 0.06 & 0.26 & n.d. (r = −0.01) & 0.38--0.39 & 0.13 \\
Discrimination (L1)* & 0.83 & 0.69 & 0.81--0.82 & 0.67--0.69 & 0.84 & 0.69--0.70 \\
Discrimination (L2)* & 0.86--0.87 & 0.70 & 0.85--0.86 & 0.68--0.70 & 0.88 & 0.70--0.72 \\
Discrimination (cosine)* & 0.97--0.98 & 0.69--0.70 & 0.97 & 0.63--0.65 & 0.98 & 0.74 \\
\bottomrule
\end{tabularx}
\end{minipage}}
\typeout{TBL S4a \the\wd\tbox\space\the\dimexpr\ht\tbox+\dp\tbox\relax\space\the\textheight}
\makebox[\textwidth][c]{\usebox{\tbox}}
\end{table}
\clearpage
\begin{table}[!htbp]
\renewcommand{\tablename}{Supplementary Table}
\renewcommand{\thetable}{S4b}
\sbox{\tbox}{\begin{minipage}{183mm}
\caption{\textbf{b, Uncorrected half-depth means.} As in a, before the Spearman--Brown correction.}
\fontsize{8}{9.5}\selectfont\setlength{\tabcolsep}{3pt}\renewcommand{\arraystretch}{1.1}
\begin{tabularx}{\linewidth}{@{}>{\raggedright\arraybackslash}Xllllll@{}}
\toprule
Metric & \begin{tabular}[b]{@{}l@{}}All: \\ random\end{tabular} & All: sgRNA & \begin{tabular}[b]{@{}l@{}}Unexpressed: \\ random\end{tabular} & \begin{tabular}[b]{@{}l@{}}Unexpressed: \\ sgRNA\end{tabular} & \begin{tabular}[b]{@{}l@{}}Knockdowns: \\ random\end{tabular} & \begin{tabular}[b]{@{}l@{}}Knockdowns: \\ sgRNA\end{tabular} \\
\midrule
DE overlap (top N) & 0.18 & 0.06 & 0.11--0.12 & 0.03 & 0.23--0.25 & 0.09 \\
DE overlap (top 50) & 0.18--0.19 & 0.06 & 0.11--0.13 & 0.03 & 0.23--0.25 & 0.09 \\
DE overlap (top 100) & 0.18 & 0.06 & 0.11--0.12 & 0.03 & 0.23--0.25 & 0.09 \\
DE overlap (top 200) & 0.18 & 0.06 & 0.11--0.12 & 0.03 & 0.23--0.25 & 0.09 \\
DE overlap (top 500) & 0.18 & 0.06 & 0.11--0.12 & 0.03 & 0.23--0.25 & 0.09 \\
DE precision (top N) & 0.18 & 0.06 & 0.11--0.12 & 0.03 & 0.24--0.25 & 0.09 \\
DE precision (top 50) & 0.18--0.19 & 0.06 & 0.11--0.13 & 0.03 & 0.25 & 0.09 \\
DE precision (top 100) & 0.18 & 0.06 & 0.11--0.12 & 0.03 & 0.24--0.25 & 0.09 \\
DE precision (top 200) & 0.17--0.18 & 0.06 & 0.11--0.12 & 0.03 & 0.24 & 0.09 \\
DE precision (top 500) & 0.18 & 0.06 & 0.11--0.12 & 0.03 & 0.24 & 0.09 \\
DE recall & 0.22--0.25 & 0.08--0.10 & 0.16--0.17 & 0.05--0.06 & 0.27--0.33 & 0.12--0.15 \\
DE direction match* & 0.72--0.74 & 0.45--0.47 & 0.68--0.69 & 0.40--0.44 & 0.75--0.79 & 0.50 \\
DE Spearman (significant genes)† & 0.67--0.69 & 0.30 & 0.67--0.69 & 0.30 & 0.67--0.69 & 0.30 \\
DE Spearman of log fold changes† & -- & -- & -- & -- & 0.26 & −0.05 \\
PR-AUC & -- & -- & -- & -- & 0.24--0.27 & 0.07--0.09 \\
ROC-AUC* & -- & -- & -- & -- & 0.72--0.76 & 0.61--0.62 \\
Pearson Δ & 0.19 & 0.03 & 0.15 & −0.01 & 0.23--0.24 & 0.07 \\
Discrimination (L1)* & 0.70--0.71 & 0.52--0.53 & 0.68--0.70 & 0.51--0.52 & 0.72--0.73 & 0.53--0.54 \\
Discrimination (L2)* & 0.76--0.77 & 0.54 & 0.74--0.75 & 0.51--0.54 & 0.78 & 0.54--0.56 \\
Discrimination (cosine)* & 0.95 & 0.53 & 0.94 & 0.46--0.48 & 0.96--0.97 & 0.59 \\
\bottomrule
\end{tabularx}
\end{minipage}}
\typeout{TBL S4b \the\wd\tbox\space\the\dimexpr\ht\tbox+\dp\tbox\relax\space\the\textheight}
\makebox[\textwidth][c]{\usebox{\tbox}}
\end{table}
\clearpage
\begin{table}[!htbp]
\renewcommand{\tablename}{Supplementary Table}
\renewcommand{\thetable}{S4c}
\sbox{\tbox}{\begin{minipage}{183mm}
\caption{\textbf{c, The seven metrics of Fig. 5b in each donor.} Spearman--Brown corrected ceilings, random halves (mean of seeds 0--2) / sgRNA halves (mean of the two directions), resting cells, selected and downsampled per donor as in a. Donors are ordered from the least to the most concentrated clones by the TRBC readout (Supplementary Table S2b). Asterisks, chance level 0.5; n.d., not defined (split-half correlation at most 0). The random-half ceiling is the higher in all 84 comparisons.}
\fontsize{7}{8.2}\selectfont\setlength{\tabcolsep}{3pt}\renewcommand{\arraystretch}{1.1}
\begin{tabularx}{\linewidth}{@{}>{\raggedright\arraybackslash}Xllll@{}}
\toprule
Metric & Donor 4 & Donor 1 & Donor 2 & Donor 3 \\
\midrule
\textbf{All perturbations} & ~ & ~ & ~ & ~ \\
DE overlap & 0.32 / 0.14 & 0.31 / 0.11 & 0.40 / 0.14 & 0.40 / 0.14 \\
DE precision & 0.33 / 0.14 & 0.30 / 0.11 & 0.40 / 0.14 & 0.39 / 0.14 \\
DE recall & 0.39 / 0.19 & 0.38 / 0.17 & 0.47 / 0.21 & 0.47 / 0.21 \\
DE direction match* & 0.80 / 0.63 & 0.84 / 0.63 & 0.91 / 0.66 & 0.92 / 0.66 \\
Pearson Δ & 0.28 / 0.08 & 0.32 / 0.06 & 0.44 / 0.08 & 0.48 / 0.09 \\
Discrimination (L1)* & 0.81 / 0.69 & 0.83 / 0.69 & 0.88 / 0.70 & 0.90 / 0.71 \\
Discrimination (cosine)* & 0.95 / 0.71 & 0.98 / 0.69 & 0.99 / 0.71 & 0.99 / 0.72 \\
\textbf{Unexpressed targets} & ~ & ~ & ~ & ~ \\
DE overlap & 0.16 / 0.03 & 0.21 / 0.05 & 0.36 / 0.11 & 0.36 / 0.12 \\
DE precision & 0.16 / 0.03 & 0.21 / 0.05 & 0.35 / 0.11 & 0.36 / 0.12 \\
DE recall & 0.23 / 0.05 & 0.29 / 0.10 & 0.43 / 0.17 & 0.45 / 0.19 \\
DE direction match* & 0.75 / 0.57 & 0.81 / 0.59 & 0.90 / 0.63 & 0.91 / 0.64 \\
Pearson Δ & 0.22 / 0.02 & 0.26 / n.d. & 0.40 / 0.03 & 0.46 / 0.04 \\
Discrimination (L1)* & 0.78 / 0.66 & 0.82 / 0.68 & 0.88 / 0.71 & 0.89 / 0.69 \\
Discrimination (cosine)* & 0.94 / 0.69 & 0.97 / 0.64 & 0.99 / 0.69 & 0.98 / 0.70 \\
\textbf{Strength-matched knockdowns} & ~ & ~ & ~ & ~ \\
DE overlap & 0.46 / 0.24 & 0.39 / 0.17 & 0.44 / 0.17 & 0.43 / 0.16 \\
DE precision & 0.47 / 0.24 & 0.39 / 0.17 & 0.44 / 0.17 & 0.43 / 0.16 \\
DE recall & 0.52 / 0.30 & 0.47 / 0.23 & 0.50 / 0.25 & 0.49 / 0.24 \\
DE direction match* & 0.85 / 0.69 & 0.87 / 0.67 & 0.92 / 0.69 & 0.92 / 0.68 \\
Pearson Δ & 0.33 / 0.14 & 0.38 / 0.13 & 0.49 / 0.13 & 0.51 / 0.14 \\
Discrimination (L1)* & 0.84 / 0.72 & 0.84 / 0.70 & 0.88 / 0.69 & 0.90 / 0.72 \\
Discrimination (cosine)* & 0.96 / 0.72 & 0.98 / 0.74 & 0.99 / 0.74 & 0.99 / 0.74 \\
Perturbations (n) & 300 & 300 & 299 & 300 \\
\bottomrule
\end{tabularx}
\end{minipage}}
\typeout{TBL S4c \the\wd\tbox\space\the\dimexpr\ht\tbox+\dp\tbox\relax\space\the\textheight}
\makebox[\textwidth][c]{\usebox{\tbox}}
\end{table}
\clearpage
\begin{table}[!htbp]
\renewcommand{\tablename}{Supplementary Table}
\renewcommand{\thetable}{S5}
\sbox{\tbox}{\begin{minipage}{183mm}
\caption{\textbf{Structure of unexpressed targets carried by the common-to-specific predictor.} For a pair of contexts X and Y, the predictor maps the common part of half A, (X\textsubscript{A} + Y\textsubscript{A})/2, to the specific part of half B, (X\textsubscript{B} − Y\textsubscript{B})/2, and gives a held-out prediction F for every perturbation. Because the specific part contains +X/2, a predictor that reads a perturbation's half-A deviation in context X passes that deviation on to F. For a set of targets S, the pass share is 2 Σ\textsubscript{S} F · X\textsubscript{B}/Σ\textsubscript{S} X\textsubscript{A} · X\textsubscript{B} (with the opposite sign for context Y): the share of the reproducible energy of the targets' own responses in X that reappears in the prediction. Responses were referenced to the mean of the unexpressed targets, and the direction of the mean of all perturbations (the template) was then projected out. The predictor used the top k principal components of the half-A
common part, ridge regression with λ chosen by inner two-fold cross-validation, and two outer folds over perturbations (seed 11). CD4 halves are random halves or sgRNA halves; in the cell-line pairs only random halves exist. Intervals are 2.5th--97.5th percentiles over 500 bootstrap resamples of targets. The shuffled control gave values of at most 0.0079 in absolute value.}
\fontsize{8}{9.5}\selectfont\setlength{\tabcolsep}{3pt}\renewcommand{\arraystretch}{1.1}
\begin{tabularx}{\linewidth}{@{}>{\raggedright\arraybackslash}Xlrlrrlr@{}}
\toprule
Context & Halves & \begin{tabular}[b]{@{}l@{}}Unexpressed, \\ n\end{tabular} & \begin{tabular}[b]{@{}l@{}}Pass, k = 20 (95\% \\ CI)\end{tabular} & k = 10 & \begin{tabular}[b]{@{}l@{}}Expressed, \\ n\end{tabular} & \begin{tabular}[b]{@{}l@{}}Pass, k = 20 \\ (95\% CI)\end{tabular} & k = 10 \\
\midrule
CD4 donor 1, resting & random & 1,671 & 0.27 (0.26--0.28) & 0.21 & 2,120 & 0.25 (0.24--0.26) & 0.19 \\
CD4 donor 1, stimulated & random & 1,671 & 0.23 (0.22--0.24) & 0.18 & 2,239 & 0.21 (0.20--0.22) & 0.16 \\
CD4 donor 1, resting & sgRNA & 1,241 & 0.00 (−0.01 to 0.01) & 0.00 & 1,424 & 0.00 (0.00--0.00) & 0.00 \\
CD4 donor 1, stimulated & sgRNA & 1,241 & 0.00 (0.00--0.00) & 0.00 & 1,511 & 0.00 (0.00--0.00) & 0.00 \\
CD4 donor 2, resting & random & 1,477 & 0.29 (0.28--0.30) & 0.23 & 1,974 & 0.27 (0.27--0.28) & 0.22 \\
CD4 donor 2, stimulated & random & 1,477 & 0.26 (0.25--0.28) & 0.22 & 1,951 & 0.26 (0.25--0.28) & 0.22 \\
CD4 donor 2, resting & sgRNA & 874 & 0.00 (−0.01 to 0.01) & 0.00 & 1,124 & 0.00 (0.00--0.00) & 0.00 \\
CD4 donor 2, stimulated & sgRNA & 874 & 0.00 (−0.03 to 0.02) & −0.01 & 1,068 & 0.00 (0.00--0.00) & 0.00 \\
K562 (K562--RPE1) & random & 181 & 0.18 (0.06--0.30) & 0.15 & 732 & 0.17 (0.13--0.20) & 0.14 \\
RPE1 (K562--RPE1) & random & 222 & 0.29 (0.25--0.33) & 0.24 & 608 & 0.38 (0.34--0.41) & 0.32 \\
Jurkat (Jurkat--HepG2) & random & 209 & 0.14 (0.10--0.17) & 0.10 & 773 & 0.23 (0.21--0.25) & 0.19 \\
HepG2 (Jurkat--HepG2) & random & 199 & 0.23 (0.15--0.31) & 0.21 & 656 & 0.32 (0.30--0.35) & 0.26 \\
\bottomrule
\end{tabularx}
\end{minipage}}
\typeout{TBL S5 \the\wd\tbox\space\the\dimexpr\ht\tbox+\dp\tbox\relax\space\the\textheight}
\makebox[\textwidth][c]{\usebox{\tbox}}
\end{table}
Under random halves in CD4, the predictor carried the structure of unexpressed targets at the same rate as that of expressed targets; under sgRNA halves it carried neither. In the four cell lines, the pass share of unexpressed targets was 0.6--1.1 times that of expressed targets. In these libraries, which were selected on a growth phenotype, the structure of unexpressed targets largely comes from the sgRNAs or target sites (Supplementary Fig. S3), so carrying it does not by itself indicate clones.

\clearpage
\begin{table}[!htbp]
\renewcommand{\tablename}{Supplementary Table}
\renewcommand{\thetable}{S6a}
\sbox{\tbox}{\begin{minipage}{183mm}
\caption{\textbf{Transfer models. a, Scores in the four donors.} Scores against the two answers, A (same sgRNA, shares clones with the input) / B (other sgRNA of the target, another transduction): Pearson correlations over the evaluation genes between predicted and answer effects, averaged over test sgRNAs, for each donor; the context mean ignores the perturbation, so its A and B are equal.}
\fontsize{7}{8.2}\selectfont\setlength{\tabcolsep}{3pt}\renewcommand{\arraystretch}{1.1}
\begin{tabularx}{\linewidth}{@{}>{\raggedright\arraybackslash}Xllll@{}}
\toprule
Model & Donor 1 & Donor 2 & Donor 3 & Donor 4 \\
\midrule
\textbf{Unexpressed targets: A / B} & ~ & ~ & ~ & ~ \\
Context mean (baseline) & 0.081 / 0.081 & 0.124 / 0.124 & 0.098 / 0.098 & 0.101 / 0.101 \\
Copy resting effect & 0.220 / −0.001 & 0.291 / 0.004 & 0.368 / 0.004 & 0.179 / 0.007 \\
Additive & 0.226 / 0.007 & 0.307 / 0.017 & 0.375 / 0.012 & 0.188 / 0.016 \\
Linear (ridge) & 0.463 / 0.020 & 0.509 / 0.011 & 0.574 / 0.039 & 0.369 / 0.051 \\
Arc perturb\_mean & 0.165 / 0.007 & 0.235 / 0.005 & 0.299 / 0.007 & 0.139 / 0.002 \\
Arc pseudobulk & 0.222 / 0.006 & 0.319 / 0.013 & 0.357 / 0.006 & 0.191 / 0.012 \\
scGen & 0.227 / −0.004 & 0.302 / 0.003 & 0.362 / 0.012 & 0.206 / 0.015 \\
STATE & 0.260 / 0.005 & 0.329 / −0.008 & 0.398 / 0.009 & 0.217 / 0.017 \\
State (SE-600M + STATE) & 0.280 / 0.014 & 0.353 / 0.002 & 0.421 / 0.012 & 0.203 / 0.014 \\
PIE (expression shift) & 0.228 / 0.108 & 0.272 / 0.101 & 0.323 / 0.100 & 0.199 / 0.093 \\
\textbf{Strength-matched knockdowns: A / B} & ~ & ~ & ~ & ~ \\
Context mean (baseline) & 0.106 / 0.106 & 0.138 / 0.138 & 0.112 / 0.112 & 0.113 / 0.113 \\
Copy resting effect & 0.228 / 0.023 & 0.302 / 0.017 & 0.359 / 0.025 & 0.189 / 0.019 \\
Additive & 0.239 / 0.034 & 0.321 / 0.034 & 0.368 / 0.034 & 0.199 / 0.029 \\
Linear (ridge) & 0.457 / 0.042 & 0.532 / 0.065 & 0.550 / 0.061 & 0.397 / 0.064 \\
Arc perturb\_mean & 0.179 / 0.026 & 0.253 / 0.017 & 0.294 / 0.021 & 0.150 / 0.017 \\
Arc pseudobulk & 0.238 / 0.030 & 0.337 / 0.040 & 0.347 / 0.042 & 0.205 / 0.030 \\
scGen & 0.230 / 0.029 & 0.300 / 0.015 & 0.354 / 0.034 & 0.210 / 0.018 \\
STATE & 0.260 / 0.020 & 0.343 / 0.034 & 0.378 / 0.030 & 0.240 / 0.030 \\
State (SE-600M + STATE) & 0.297 / 0.038 & 0.389 / 0.030 & 0.394 / 0.033 & 0.228 / 0.026 \\
PIE (expression shift) & 0.233 / 0.122 & 0.295 / 0.119 & 0.332 / 0.118 & 0.210 / 0.109 \\
Test sgRNAs, unexpressed / knockdowns (n) & 254 / 246 & 270 / 252 & 254 / 252 & 280 / 274 \\
\bottomrule
\end{tabularx}
\end{minipage}}
\typeout{TBL S6a \the\wd\tbox\space\the\dimexpr\ht\tbox+\dp\tbox\relax\space\the\textheight}
\makebox[\textwidth][c]{\usebox{\tbox}}
\end{table}
\clearpage
\begin{table}[!htbp]
\renewcommand{\tablename}{Supplementary Table}
\renewcommand{\thetable}{S6b}
\sbox{\tbox}{\begin{minipage}{183mm}
\caption{\textbf{b, Scores against the two answers in CD4 donor 1, with standard errors.} scGen is in a. A, answer from the same sgRNA (shares clones with the input); B, answer from the other sgRNA against the same target (another transduction). Scores are Pearson correlations over the evaluation genes between predicted and answer effects, averaged over test sgRNAs; standard errors of A − B are clustered by target. For reference, the stimulated effects of the two sgRNAs of a target agree at r = 0.041 (unexpressed targets) and 0.059 (knockdowns).}
\fontsize{8}{9.5}\selectfont\setlength{\tabcolsep}{3pt}\renewcommand{\arraystretch}{1.1}
\begin{tabularx}{\linewidth}{@{}>{\raggedright\arraybackslash}Xrrlrrl>{\raggedright\arraybackslash}X@{}}
\toprule
Model & \begin{tabular}[b]{@{}l@{}}Unexpressed: \\ A\end{tabular} & \begin{tabular}[b]{@{}l@{}}Unexpressed: \\ B\end{tabular} & \begin{tabular}[b]{@{}l@{}}Unexpressed: A \\ − B (SE)\end{tabular} & \begin{tabular}[b]{@{}l@{}}Knockdowns: \\ A\end{tabular} & \begin{tabular}[b]{@{}l@{}}Knockdowns: \\ B\end{tabular} & \begin{tabular}[b]{@{}l@{}}Knockdowns: A \\ − B (SE)\end{tabular} & Note \\
\midrule
Context mean & 0.081 & 0.081 & 0.000 (0.012) & 0.106 & 0.106 & 0.000 (0.012) & ignores the perturbation \\
Copy resting effect & 0.220 & −0.001 & 0.221 (0.008) & 0.228 & 0.023 & 0.205 (0.009) & ~ \\
Additive & 0.226 & 0.007 & 0.220 (0.008) & 0.239 & 0.034 & 0.204 (0.009) & ~ \\
Linear (ridge) & 0.463 & 0.020 & 0.443 (0.020) & 0.457 & 0.042 & 0.415 (0.020) & ~ \\
Arc perturb\_mean & 0.165 & 0.007 & 0.158 (0.009) & 0.179 & 0.026 & 0.153 (0.010) & ~ \\
Arc pseudobulk & 0.222 & 0.006 & 0.216 (0.012) & 0.238 & 0.030 & 0.208 (0.014) & ~ \\
STATE & 0.260 & 0.005 & 0.254 (0.016) & 0.260 & 0.020 & 0.241 (0.016) & ~ \\
State (SE-600M + STATE) & 0.280 & 0.014 & 0.266 (0.016) & 0.297 & 0.038 & 0.259 (0.018) & ~ \\
PIE (expression shift) & 0.228 & 0.108 & 0.120 (0.015) & 0.233 & 0.122 & 0.111 (0.018) & ~ \\
PIE (log2 fold change) & 0.026 & 0.011 & 0.015 (0.003) & 0.029 & 0.014 & 0.015 (0.003) & ~ \\
Stack (zero-shot) & 0.038 & 0.003 & 0.036 (0.014) & 0.023 & −0.005 & 0.028 (0.015) & 1,672 of 2,152 genes \\
biolord (adapted) & 0.030 & 0.004 & 0.027 (0.026) & 0.020 & −0.033 & 0.053 (0.028) & fit r = 0.35 \\
biolord (tutorial) & 0.005 & 0.010 & −0.006 (0.022) & −0.035 & −0.031 & −0.005 (0.023) & did not learn sgRNAs \\
CPA (default) & −0.001 & 0.003 & −0.003 (0.017) & −0.031 & 0.000 & −0.031 (0.020) & did not learn sgRNAs \\
CPA (adapted) & 0.001 & −0.002 & 0.003 (0.017) & −0.021 & −0.003 & −0.018 (0.020) & did not learn sgRNAs \\
\bottomrule
\end{tabularx}
\end{minipage}}
\typeout{TBL S6b \the\wd\tbox\space\the\dimexpr\ht\tbox+\dp\tbox\relax\space\the\textheight}
\makebox[\textwidth][c]{\usebox{\tbox}}
\end{table}
\textbf{Settings.} Context mean, copy, additive and linear are our baselines: the mean stimulated effect of the training sgRNAs; the sgRNA's resting effect; the resting effect plus the mean change between states of the training sgRNAs; and a ridge regression from 50 principal components of the resting effect to the stimulated effect, with λ chosen by five-fold cross-validation over training sgRNAs. STATE, Arc pseudobulk and Arc perturb\_mean were trained with arc-state 0.11.1 (cell-load 0.10.4; \texttt{state\ tx\ train}) with the settings of the package's README example (40,000 steps, batch size 8; for STATE, a transformer backbone of 8 layers with hidden size 768 and the energy loss), on log1p CP10K expression of about 2,150 evaluation genes rather than the authors' highly variable genes, with the sgRNA as the perturbation and the state as the cell type, the few-shot split described in Methods, and the stimulated cells of 40 training sgRNAs for validation; predictions used
\texttt{state\ tx\ predict\ -\/-predict-only}, and effects were taken relative to the predicted non-targeting stimulated cells. State used the SE-600M foundation model to embed cells before STATE. PIE\textsuperscript{36} (arc-pie 1.0.0) used the authors' setting for the Replogle data (5,000 steps, effective batch size 128), with the released knowledge sources except ncbi\_text, and the checkpoint with the best validation AUPRC. scGen\textsuperscript{35} (version 2.1.1 from its GitHub repository, with scvi-tools 0.20.3) was trained with the settings of its perturbation-prediction tutorial (at most 100 epochs, batch size 32, early stopping with patience 25), with the sgRNA as its cell type and the state as its condition, and the stimulated cells of the test sgRNAs held out; prediction follows the package's predict function: the latent shift between states, computed as the difference of equal-weight means over sgRNAs (the package's balancing, computed exactly instead of by resampling), is
added to the latent representation of each test sgRNA's resting cells, which are then decoded. Stack\textsuperscript{37} (Stack-Large-Aligned) was run zero-shot as in its tutorial, with the resting cells of each test sgRNA as context and 1,000 stimulated control cells as target. CPA\textsuperscript{38} and biolord\textsuperscript{39} were run with their default or tutorial settings and with adapted settings. \textbf{Fit checks.} Re-predicting 200 training sgRNAs as if they were test sgRNAs, STATE correlated with their observed stimulated effects at r = 0.43--0.52 across the four donors, against −0.001 to 0.012 with other training sgRNAs, so it learned per-sgRNA effects; CPA and biolord with tutorial settings had fit correlations of at most 0.025 and did not learn per-sgRNA effects, so their A − B cannot be judged. All models except Stack, CPA and biolord were run in all four donors (a); in donors 2--4, Arc pseudobulk, State and PIE were run on another server with the same code,
environments and settings.

\clearpage
\begin{table}[!htbp]
\renewcommand{\tablename}{Supplementary Table}
\renewcommand{\thetable}{S7}
\sbox{\tbox}{\begin{minipage}{183mm}
\caption{\textbf{Do the contexts of public multi-context screens come from one transduction?} Contexts from one transduction (a transduced, selected and expanded population later split, treated, differentiated or sampled) share clones; contexts transduced separately (other cell lines, donors or screens) do not. Calls rest on the transduction and screen-design paragraphs of each paper's methods, read in full text; the quoted sentences and references are in notes/multicontext\_transduction\_survey.md of the code repository. MOI, multiplicity of infection, given where the cell number is not stated. Benchmark use was checked in the benchmark papers or repositories. Sharing clones is necessary for the inflation of transfer scores; its size also depends on the effective number of clones per perturbation.}
\fontsize{8}{9.5}\selectfont\setlength{\tabcolsep}{3pt}\renewcommand{\arraystretch}{1.1}
\begin{tabularx}{\linewidth}{@{}>{\raggedright\arraybackslash}X>{\raggedright\arraybackslash}X>{\raggedright\arraybackslash}Xl>{\raggedright\arraybackslash}X@{}}
\toprule
Data set & Contexts & Contexts from & \begin{tabular}[b]{@{}l@{}}Transduced \\ cells per \\ sgRNA\end{tabular} & Use in benchmarks \\
\midrule
Zhu et al.~2026, primary CD4+ T cells & rest, stimulated 8 h and 48 h; 4 donors & one transduction per donor; separate across donors & MOI about 0.2 & this study \\
Frangieh et al.~2021, melanoma & control, IFN-γ, T-cell co-culture & one transduction & about 1,000 & PerturBench (transfer across conditions); scPerturb \\
Papalexi et al.~2021, THP-1 & stimulated, unstimulated & one transduction per replicate & not stated & scPerturb; scPerturbBench \\
Datlinger et al.~2017, Jurkat & stimulated, unstimulated & one transduction & not stated & scPerturb; scPerturbBench \\
Dixit et al.~2016, dendritic cells and K562 & 0 h and 3 h after LPS; day 7 and day 13 & one transduction per pool & not stated & scPerturb \\
Tian et al.~2019, iPSC and neurons & iPSC, day-7 neuron & one transduction & not stated & scPerturb \\
Schmidt et al.~2022, primary T cells & stimulated, unstimulated; 2 donors & one transduction per donor & MOI 0.3 & none found \\
Shifrut et al.~2018, primary T cells & stimulated, resting; 2 donors & one transduction per donor (resting arm not described) & not stated & scPerturb \\
McFaline-Figueroa et al.~2024, glioblastoma lines & drugs and doses; 3 cell lines & one transduction per line; separate across lines & MOI about 0.1 & PerturBench \\
Zhou et al.~2023, mouse T cells & input, tumour & one transduction & not stated & none found \\
Replogle et al.~2022, K562 and RPE1 & 3 screens & separate transductions & about 1,000 & State; PIE; scPerturb; scPerturbBench \\
Nadig et al.~2025, Jurkat and HepG2 & 2 screens & separate transductions & about 1,000 & State (held-out cell line); PIE \\
Jiang et al.~2025, 6 cell lines & 5 cytokines per line & separate transduction per line & not stated & PerturBench \\
Tian et al.~2021, neurons & CRISPRi, CRISPRa & separate transductions & MOI 0.1--0.2 & scPerturb; scPerturbBench \\
\bottomrule
\end{tabularx}
\end{minipage}}
\typeout{TBL S7 \the\wd\tbox\space\the\dimexpr\ht\tbox+\dp\tbox\relax\space\the\textheight}
\makebox[\textwidth][c]{\usebox{\tbox}}
\end{table}
\clearpage
\begin{table}[!htbp]
\renewcommand{\tablename}{Supplementary Table}
\renewcommand{\thetable}{S8}
\sbox{\tbox}{\begin{minipage}{183mm}
\caption{\textbf{Data sets, their uses and accessions.}}
\fontsize{8}{9.5}\selectfont\setlength{\tabcolsep}{3pt}\renewcommand{\arraystretch}{1.1}
\begin{tabularx}{\linewidth}{@{}>{\raggedright\arraybackslash}X>{\raggedright\arraybackslash}X>{\raggedright\arraybackslash}X>{\raggedright\arraybackslash}X@{}}
\toprule
Data & System & Use & Source \\
\midrule
Zhu et al. & primary human CD4+ T cells, CRISPRi, genome-wide, four donors & main analyses & authors' data release (S3, marson2025\_data) \\
Replogle et al. & K562 (essential, genome-wide) and RPE1, CRISPRi & null controls, clone strength, predictor & figshare+ 20029387 \\
Nadig et al. & HepG2 and Jurkat, CRISPRi, essential genes & null controls, clone strength & GEO GSE264667 \\
Virtual Cell Challenge 2025 & H1 hESC, CRISPRi & null controls, clone strength & public training data, Arc Virtual Cell Atlas \\
Li et al. & human brain organoids, clone barcodes & clone halves, test of the relation & Zenodo 7083558; ArrayExpress E-MTAB-13148 \\
Shi et al. & mouse whole brain, AAV after birth, single nuclei & clone strength, reference within mouse & Hugging Face perturbai/wholebrain\_crispr\_atlas \\
Jin et al. & embryonic mouse cortex, lentivirus in utero & clone strength (library halves) & GEO GSE157977 \\
Knudsen et al. & human CAR-T cells in NSG mice & clone strength & GEO GSE302504 \\
Fagerberg et al. & mouse P14 CD8 T cells in LCMV-infected mice & clone strength & FITdb, Zenodo 22212134 \\
Zhou et al. & mouse OT-I CD8 T cells in tumour-bearing mice & clone strength & GEO GSE216800 \\
Schmidt et al. & primary human T cells, CRISPRa, two donors & clone strength & FITdb, Zenodo 22212134 \\
Cao et al. & primary human AML cells, in vitro & clone strength & GEO GSE221578 \\
Frangieh et al. & melanoma cell line, three treatments & clone strength, transfer models & scPerturb, Zenodo 13350497 \\
\bottomrule
\end{tabularx}
\end{minipage}}
\typeout{TBL S8 \the\wd\tbox\space\the\dimexpr\ht\tbox+\dp\tbox\relax\space\the\textheight}
\makebox[\textwidth][c]{\usebox{\tbox}}
\end{table}
\clearpage
\begin{figure}[!htbp]
\renewcommand{\figurename}{Supplementary Figure}\renewcommand{\thefigure}{S1}
\makebox[\textwidth][c]{\includegraphics[width=170mm]{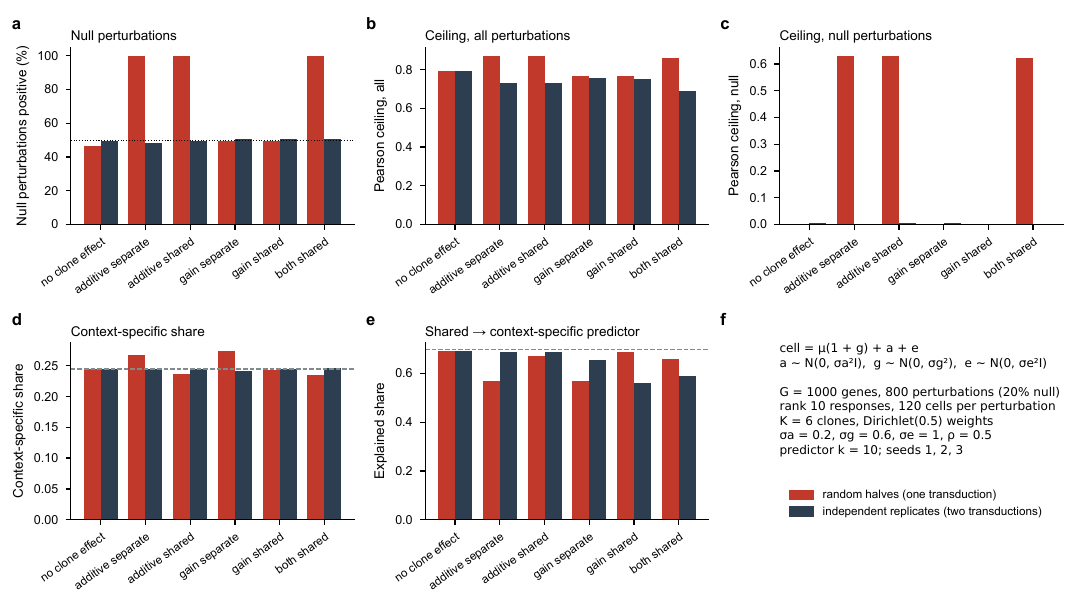}}
\caption{\textbf{Three-term model: expectations and simulation scenarios.} Simulated screens with two contexts, under random halves (red; both halves from one transduction) and independent replicates (blue; each half from its own transduction), for six scenarios: no clone effect; additive clone states only, with clones separate between contexts or shared by them; multiplicative clone gains only, separate or shared; and both terms, shared. \textbf{a}, Percentage of null perturbations with a positive cross-half signal; dotted line, 50\%. \textbf{b}, \textbf{c}, Spearman--Brown corrected Pearson ceiling for all perturbations and for null perturbations. \textbf{d}, Context-specific share; dashed line, true value. \textbf{e}, Explained share of the predictor of the specific part from the common part (labelled shared-to-specific in the figure); dashed line, true value. \textbf{f}, Model and parameters. Additive clone states make null perturbations reproducible and raise both ceilings, under random
halves only. Gains leave null perturbations unaffected. The context-specific share changes little in any scenario. Clone effects that are separate between contexts lower the explained share under random halves, and shared gains lower it across independent replicates. The scenarios without and with true structure are shown in Fig. 6e,f.~Means over three seeds.}
\end{figure}

\clearpage
\begin{figure}[!htbp]
\renewcommand{\figurename}{Supplementary Figure}\renewcommand{\thefigure}{S2}
\makebox[\textwidth][c]{\includegraphics[width=89mm]{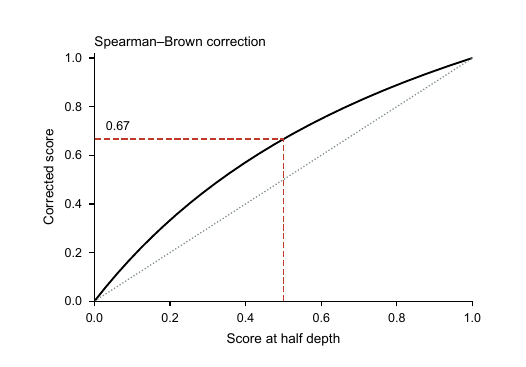}}
\caption{\textbf{The Spearman--Brown correction at chance level.} The correction 2r/(1 + r) extrapolates a score computed between two halves to the full number of cells. For metrics whose chance level is 0.5, such as direction match and discrimination, it maps chance to 0.67 (dashed lines), so a corrected ceiling near 0.67 can mean no reproducible signal. Applying the correction to the score above chance would avoid this. Dotted line, identity.}
\end{figure}

\clearpage
\begin{figure}[!htbp]
\renewcommand{\figurename}{Supplementary Figure}\renewcommand{\thefigure}{S3}
\makebox[\textwidth][c]{\includegraphics[width=170mm]{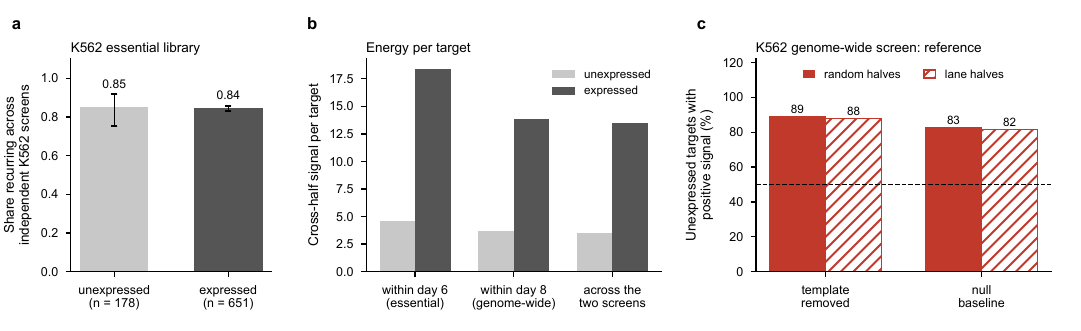}}
\caption{\textbf{Unexpressed targets in K562 screens.} \textbf{a}, Share of the within-screen structure of unexpressed targets (n = 178) and expressed targets (n = 651) of the K562 essential-gene library that recurs in an independent K562 screen (genome-wide screen, part 1): summed cross-screen signal divided by the geometric mean of the summed within-screen signals. Error bars, bootstrap 95\% intervals over targets (1,000 resamples). Unexpressed targets recur as much as expressed ones, so their structure comes mainly from the sgRNAs or target sites, not from clones. \textbf{b}, Cross-half signal per target within the day-6 essential-gene screen, within the day-8 genome-wide screen and across the two screens. The non-recurring part of the null signal, (1 − 0.85) × 4.6, is at most about 4\% of the expressed signal of 18.4 per target; because it also contains differences between the two screens, it is an upper bound on the clonal part. \textbf{c}, K562 genome-wide screen (part 1): share of
unexpressed targets with a positive cross-half signal when responses are referenced to the mean of all perturbations (template removed) or to the mean of the unexpressed targets (null baseline), for random halves and for halves formed by sequencing lane. Template removal turns each null perturbation into a reproducible negative template and raises the share. Dashed line, 50\%.}
\end{figure}

\end{document}